%% file: main.tex
\documentclass[pre,superscriptaddress,tightenlines,onecolumn,showpacs,longbibliography,preprint,floatfix]{revtex4-1}

\pdfoutput=1

\usepackage{amsmath,amsfonts,amssymb,bm,empheq,latexsym}
\usepackage{subfigure}

\usepackage{ulem}
\usepackage{tensor}
\usepackage{blkarray}
\usepackage{mathtools}
\usepackage{multirow}

\usepackage{color}
\usepackage[dvipsnames]{xcolor}
\definecolor{nblue}{RGB}{28,130,185}

\definecolor{cgreen}{RGB}{76,153,0}

\definecolor{myorange}{RGB}{245,156,74}

\usepackage{hyperref}
\hypersetup{
  colorlinks=true,
  citecolor=magenta,
  urlcolor=-myorange
}

\newcommand{\ddt}[1]{\mathrm{d}#1/\mathrm{d}t}
\newcommand{\ddthat}[1]{\mathrm{d}#1/\mathrm{d}\hat t}

\newcommand{\dd}{\mathrm{d}}

\newcommand{\ddroit}[2]{\frac{\mathrm{d}#1}{\mathrm{d}#2}}
\newcommand{\drond}[2]{\frac{\partial #1}{\partial #2}}

\begin{document}

\title{Fluid pumping and active flexoelectricity can promote lumen nucleation in cell assemblies}

\input{authors.tex}

\date{\today}
\input{abstract.tex}

\maketitle

\input{introduction.tex}
\input{text.tex}
\input{conclusion.tex}

\input{acknow.tex}

\clearpage
\appendix

\input{appendices.tex}

\clearpage
\bibliography{biblio.bib} 

\end{document}

%% file: authors.tex
\author{Charlie Duclut}
\affiliation{Max-Planck-Institut f\"ur Physik komplexer Systeme, N\"othnitzer Str.~38, 01187 Dresden, Germany}

\author{Niladri Sarkar}
\affiliation{Laboratoire Physico Chimie Curie, UMR 168, Institut Curie, PSL Research University, CNRS, Sorbonne Universit\'e, 75005 Paris, France}%
\affiliation{Instituut-Lorentz, Universiteit Leiden, P.O. Box 9506, 2300 RA Leiden, The Netherlands}
 
\author{Jacques Prost}
\affiliation{Laboratoire Physico Chimie Curie, UMR 168, Institut Curie, PSL Research University, CNRS, Sorbonne Universit\'e, 75005 Paris, France}
\affiliation{Mechanobiology Institute, National University of Singapore, 117411 Singapore}

\author{Frank J\"ulicher}\email{julicher@pks.mpg.de}
\affiliation{Max-Planck-Institut f\"ur Physik komplexer Systeme, N\"othnitzer Str.~38, 01187 Dresden, Germany}%
\affiliation{Center for Systems Biology Dresden, Pfotenhauerstr. 108, 01307 Dresden, Germany}
\affiliation{Cluster of Excellence Physics of Life, TU Dresden, 01062 Dresden, Germany}

%% file: abstract.tex
\begin{abstract}

We discuss the physical mechanisms that promote or suppress the  nucleation of a fluid-filled lumen inside a cell assembly or a tissue. We discuss lumen formation in a continuum theory of tissue material properties in which the tissue is described as a two-fluid system to account for its permeation by the interstitial fluid, and we include fluid pumping as well as active electric effects. Considering a spherical geometry and a polarized tissue, our work shows that fluid pumping and tissue flexoelectricity play a crucial role in lumen formation. We furthermore explore the large variety of long-time states that are accessible for the cell aggregate and its lumen. Our work reveals a role of the coupling of mechanical, electrical and hydraulic phenomena in tissue lumen formation.
    
\end{abstract}

%% file: introduction.tex
\section*{Introduction}

A fundamental problem in biology is to understand the collective organization of many cells that can give rise to complex structures and morphologies. Such phenomena can be studied either in living embryos or developing organisms, but also \textit{in vitro} for example in organoid systems that recapitulate morphogenetic processes~\cite{lancaster2014,simunovic2017} or by studying even simpler cell assemblies. In these systems, it is often observed that liquid-filled cavities or lumens appear within cell assemblies~\cite{sigurbjornsdottir2014}: in respiratory, circulatory, and secretory organs, it is typically an interconnected network of tubular lumens that forms~\cite{andrew2010}. Alternatively, spherical lumens can also form, such as cysts, acini, alveoli or follicles in mammalian epithelial organs~\cite{obrien2002,ferrari2008}. Strikingly, this ability of cell assemblies to self-organize and form internal fluid cavities is maintained 
in simpler model systems such as organoids and even in multicellular spheroids formed by a single cell type. Examples include Madin--Darby canine kidney cells and mammary epithelial cells (MCF-10A) that are observed to form polarized spherical aggregates with liquid-filled lumen~\cite{martin-belmonte2008,debnath2002}. 

The formation of a well-delimited cavity surrounded by a cohesive cellular structure has been observed to rely on various mechanisms~\cite{lubarsky2003,andrew2010,martin-belmonte2008,debnath2002}. Programmed cell death induced at the structure center for instance leads to lumen formation by a process known as cavitation~\cite{lubarsky2003,sigurbjornsdottir2014}. In addition, lumenogenesis must also rely on cells ability to transport water and ions in a collective fashion to open fluid-filled cavities. This capacity of cells to pump fluid has for instance been quantified in experiments on rabbit corneal epithelia~\cite{maurice1972,sanchez2002}. The ion pumps that are necessary to generate fluid flows also produce ion flows that can lead to the build up of an electric field across the tissue. Strong experimental evidence indeed supports the presence of a voltage difference across many tissues~\cite{cereijido1978,josephson1979,hay1985}. More generally, control of cell proliferation by electric mechanisms 
has received an increasing attention~\cite{blackiston2009,cervera2018,levin2018}, and is for instance suspected to play a role in zebrafish fin growth control~\cite{daane2018}.

In this paper, we use a continuum theory of radially polarized cell spheroids to reveal key physical mechanisms underlying lumen formation. This coarse-grained approach is especially suited to study the combined effects of fluid permeation, electric fields and currents, as well as mechanical stresses stemming from cell division and death~\cite{ranft2012,sarkar2019}. We show that lumen formation is an active nucleation problem, governed by tissue growth, fluid pumping and active electric effects. In particular, we discover a surprising role of tissue flexoelectricity in lumen nucleation. Flexoelectricity was first observed  as a bending of the nematic order of liquid crystals when an external electric field is applied~\cite{meyer1969,degennes1974}. This also implies that electric fields are generated when liquid crystal orientational order is bent. In tissues, flexoelectricity describes the emergence of electric fields when cell polarity orientation is bent or deformed. 
We also discuss the state diagram of lumen formation as a function of key parameters and explore the interplay of lumen and spheroid growth at long time.

The structure of this manuscript is as follows. We first present the geometry, boundary conditions and material properties of a permeated tissue in the presence of electric fields. Having derived the dynamical equations for the inner and outer radii of the spheroid, we then focus on lumen nucleation and we show how lumen formation at early time is influenced by pumping and active electric effects. Finally, we explore in Sec.~\ref{sec_spheroidFate} the long-time states of the spheroid and its lumen.

%% file: text.tex
\section{Constitutive equations of a permeated tissue in the presence of electric fields}

        \subsection{Tissue geometry and notation}

Following Ref.~\cite{sarkar2019}, we adopt in this paper a coarse-grained, hydrodynamic description of tissues to study the formation of lumen in a spherical aggregate of cells. In the simplified model we consider here, the tissue is permeated by the interstitial fluid and described in a two-fluid framework. The cell active pumping as well as the electric field and current are introduced in the constitutive equations that describe the tissue properties.

\begin{figure}[t]
	\centering 
	\null \hfill
    {\includegraphics[width=0.5\linewidth]{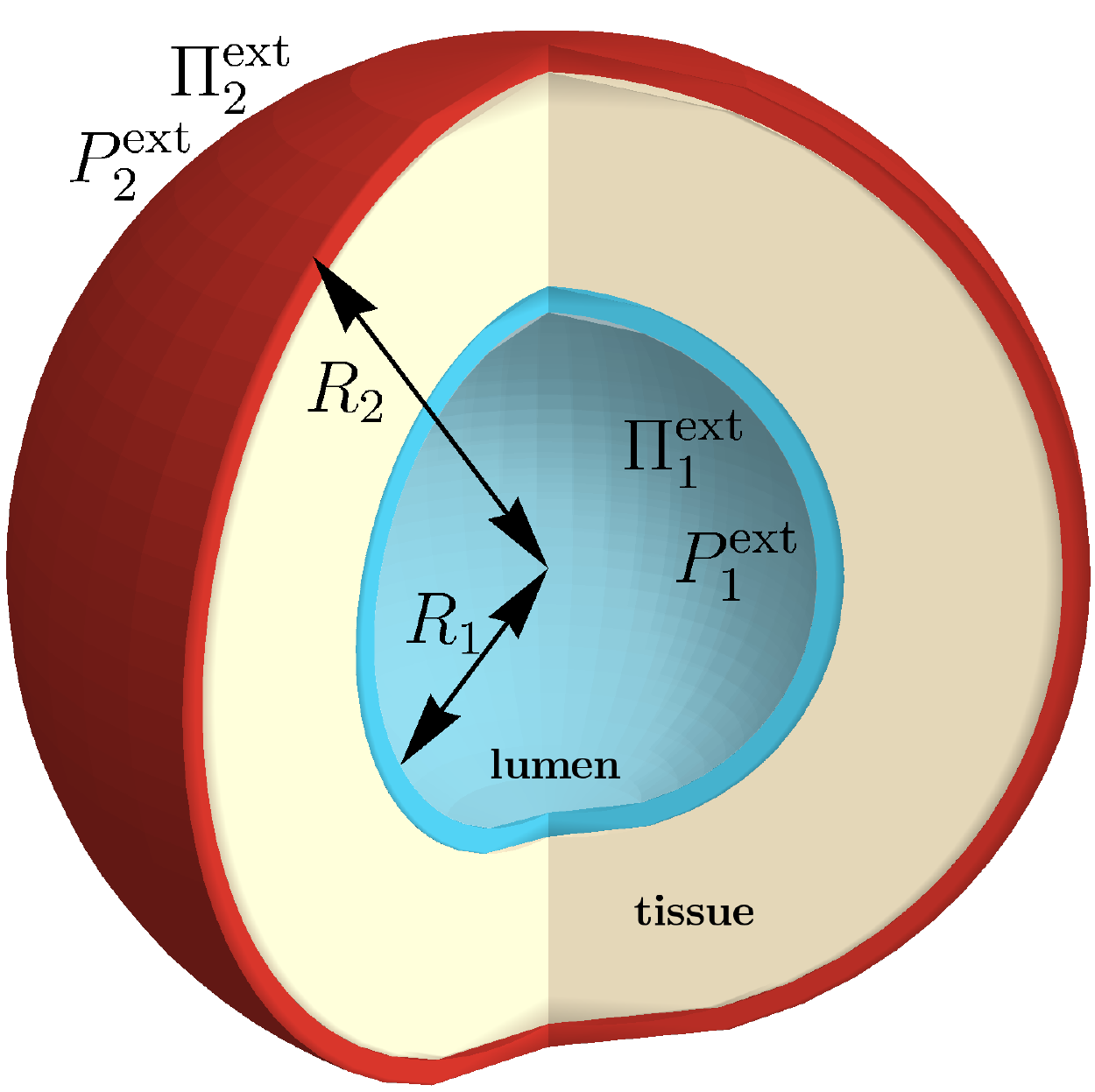}}
    \hfill \null
	\caption{Sketch of the model used for a spheroid enclosing a lumen: the spherical cell aggregate of radius $R_2$ encloses a spherical cavity of radius $R_1$ filled with fluid, the lumen. In the lumen (resp. outside the spheroid), we denote the fluid pressure as $P_1^{\rm ext}$ (resp. $P_2^{\rm ext}$) and the osmotic pressure of ions that cannot enter the tissue as $\Pi_1^{\rm ext}$ (resp. $\Pi_2^{\rm ext}$). The inner cyan (resp. outer red) shell indicates a small volume of the tissue close to the lumen (resp. close to the outside) with a cell division rate different from the bulk division rate. Fluid exchange driven by osmotic conditions is also happening at the inner and outer boundaries, see main text for details on the boundary conditions.}
    \label{fig_notation}
\end{figure}

Although spheroids and lumens are found with a large variety of shapes, we consider here a simplified geometry with a spherical symmetry, such that the dynamical quantities only vary along the radial direction $\bm e_r$. The spheroid has an outer radius $R_2$ and encloses a fluid-filled lumen of radius $R_1<R_2$, as illustrated in Fig.~\ref{fig_notation}. The spherical aggregate is surrounded on the outside (for $r>R_2$) by a fluid with a hydrostatic pressure $P^{\mathrm{ext}}_2$, and containing osmolites that enter neither the tissue nor the interstitial fluid, such that there exists an osmotic pressure $\Pi^{\mathrm{ext}}_2$. We define similarly the hydrostatic and osmotic pressures $P^{\mathrm{ext}}_1$ and $\Pi^{\mathrm{ext}}_1$ inside the lumen ($r<R_1$). The precise boundary conditions involving in particular tissue permeation at the inner and outer surfaces as well as growth rates at the boundaries will be specified later, and we now focus on the bulk equations describing the cell aggregate.

   \subsection{Permeation of the spheroid by the interstitial fluid}
   
To account for fluid transport and the formation of a fluid-filled cavity inside the spheroid, we decompose the total tissue stress as $\sigma_{\alpha\beta}=\sigma^{\rm c}_{\alpha\beta}
+ \sigma^{\rm f}_{\alpha\beta}$, where $\sigma^{\rm c}_{\alpha\beta}$ is the stress associated with the cells and $\sigma^{\rm f}_{\alpha\beta}$ is the stress associated with the interstitial fluid. For simplicity, we moreover consider that the anisotropic stress in the interstitial fluid vanishes over length scales large compared to that of the cells. In this case the interstitial fluid flow is driven by pressure gradient and the fluid stress is simply $\sigma^{\rm f}_{\alpha\beta}=-P^{\rm f}\delta_{\alpha\beta}$~\cite{ranft2012}. Neglecting inertia and in the absence of external forces, the force balance in spherical coordinates reads:
\begin{align}
    \frac{1}{r^2} \drond{}{r}\left(r^2\tilde\sigma^{\rm c}_{rr}\right)-\frac{\tilde\sigma^{\rm c}_{\theta\theta}+\tilde\sigma^{\rm c}_{\varphi\varphi}}{r}+\drond{\sigma^{\rm c}}{r} = \drond{P^{\rm f}}{r} \, ,
\label{eq_forceBalance_spherical}
\end{align}
where we have decomposed the cell stress as the sum of an isotropic and an anisotropic part: $\sigma^{\rm c}_{\alpha\beta}=\sigma^{\rm c}\delta_{\alpha\beta}+\tilde\sigma^{\rm c}_{\alpha\beta}$, and where $\tilde\sigma^{\rm c}_{\theta\theta} =\tilde\sigma^{\rm c}_{\varphi\varphi}=-\tilde\sigma^{\rm c}_{rr}/2$ because of the spherical symmetry of the cell aggregate.

   \subsection{Isotropic and anisotropic cell stress in a permeated cell aggregate}
    
We now introduce the constitutive equations for a polar tissue permeated by a fluid and subject to an electric field due to ion transport. Cells are considered polar, i.e., they exhibit a structural anisotropy that can be characterized by a unit polarity vector ${\bm p}$. This introduces a nematic order parameter for the cells given by $q_{\alpha\beta}=p_\alpha p_\beta -\delta_{\alpha\beta}/3$. Notice that we consider in the following that cells are polarized along the radial direction, such that $\bm p= \bm e_r$.

Even though tissues are elastic at shorter time scales, cell division and apoptosis enable them to release stress at longer time scales and to become effectively fluid-like~\cite{ranft2010,ranft2012}. This fluidization of the cells is made possible because cells can probe and sense the local stress and react to it, either by dividing or starting apoptosis, or by generating active stress (for instance in the cytoskeleton) as a result of energy consumption at the molecular level.

An expansion based on symmetry near the homeostatic pressure $P^{\rm c}_{\rm h}$ -- defined as the pressure at which cell death exactly compensates cell division in the isotropic state~\cite{basan2009} -- allows us to derive a constitutive equation for the isotropic cell stress $\sigma^{\rm c}$ (see Ref.~\cite{sarkar2019} for more details). In the quasistatic limit, the elastic stress has relaxed due to cell division and death, and the constitutive equation for the isotropic part reads\footnote{Notice that we have added the term $\nu_2$ in Eq.~\eqref{eq_sigma} and the term $\nu_4$ in Eq.~\eqref{eq_sigmaTilde} that were not present in Ref.~\cite{sarkar2019} although they are allowed by symmetry.}:
\begin{align}
\sigma^{\rm c} + P^{\rm c}_{\rm h} = \bar\eta 
v^{\rm c}_{\gamma\gamma} - \nu_0 \tilde\sigma^{\rm c}_{\alpha\beta}q_{\alpha\beta} - \nu_1 p_\alpha E_\alpha  - \nu_2 p_\alpha (v^{\rm c}_\alpha-v^{\rm f}_\alpha) \, , \label{eq_sigma}
\end{align}
where we have introduced the cell and fluid velocities, $v^{\rm c}_\alpha$ and $v^{\rm f}_\alpha$. We have also defined the cell strain rate tensor $v^{\rm c}_{\alpha\beta}=(\partial_\alpha v^{\rm c}_\beta+\partial_\beta v^{\rm c}_\alpha)/2$ and the summation over repeated indices is implied. We have introduced $\bar\eta$ the effective bulk viscosity of the tissue due to the response of cell growth to stress, $\nu_0$ is a dimensionless coefficient that takes into account the possible dependence of the homeostatic pressure on the anisotropic part of the stress, $\nu_1$ characterizes the influence of the electric field on the homeostatic pressure, and $\nu_2$ is a coefficient accounting for the effects of the relative motion of the cells and the interstitial fluid to homeostatic pressure~\cite{sarkar2019}.

A similar expansion for the traceless anisotropic part of the stress tensor reads:
\begin{align}
\tilde\sigma^{\rm c}_{\alpha\beta}=2\eta \tilde v^{\rm c}_{\alpha\beta}+\zeta q_{\alpha\beta}-\nu_3 [E_\alpha p_\beta]_{\rm st} -\nu_4 [v_\alpha p_\beta]_{\rm st} \, . \label{eq_sigmaTilde}
\end{align}
where $\tilde v^{\rm c}_{\alpha\beta}$ is the traceless part of the cell strain rate tensor, and we have moreover defined the symmetric traceless part of the projection of the cell polarity on the electric field: $[E_\alpha p_\beta]_{\rm st}\equiv p_\alpha E_\beta + p_\beta E_\alpha - (2/3) p_\gamma E_\gamma \delta_{\alpha\beta}$ and similarly for the projection of the cell polarity on the velocity difference: $[v_\alpha p_\beta]_{\rm st} \equiv p_\alpha (v^{\rm c}_\beta-v^{\rm f}_\beta)  + p_\beta (v^{\rm c}_\alpha-v^{\rm f}_\alpha) - (2/3) p_\gamma (v^{\rm c}_\gamma-v^{\rm f}_\gamma)  \delta_{\alpha\beta}$. We introduce $\eta$ the isotropic shear viscosity of the tissue, and we ignore the fourth rank tensor nature of the viscosity. The coefficient $\nu_3$ describes the coupling of the electric field to the anisotropic cell stress and $\nu_4$ represents the magnitude of the coupling induced by the interstitial fluid flow through the anisotropic cells. The magnitude of the active anisotropic cell stress $\zeta$ can include both a collective component arising from cell division and death, and a contribution from each cell due to the activity of their cytoskeleton~\cite{delarue2014}. This active stress can be regulated by the cells and we therefore consider that it depends on the local pressure at linear order as:
\begin{align}
    \zeta = \zeta_0 - \zeta_1 (\sigma^{\rm c}+P^{\rm c}_{\rm h}) \, ,
    \label{eq_activeStress}
\end{align}
where $\zeta_{0,1}$ are assumed to be constant. Notice that $\zeta_1$ is a dimensionless parameter. 

    \subsection{Fluid permeation and electric currents}

To close our system of equations, we moreover write a constitutive equation for the momentum exchange $f_\alpha = \partial_\beta \sigma^{\rm f}_{\alpha\beta}$ between interstitial fluid and cells~\cite{ranft2012}. This term, which is balanced by the interstitial fluid pressure gradient $-\partial_r P^{\rm f}$ in our geometry, can be expressed as:
\begin{align}
f_\alpha\!=\!-\kappa(v^{\rm c}_\alpha \!-\! v^{\rm f}_\alpha) + \lambda_1 p_\alpha \!+ \lambda_2 E_\alpha \!+ \lambda_3 q_{\alpha\beta}E_\beta \!+ \lambda_4 \partial_\beta q_{\alpha\beta} \, .
\label{eq_momentum}
\end{align}
The coefficient $\kappa$ describes the friction due to the (relative) flow of the interstitial fluid in the nanometric cleft between cells, and corresponds to Darcy's law~\cite{darcy1856} in the description of porous media. The second term on the right-hand side represents the active pumping of fluid by the cells, with $\lambda_1$ the active pumping coefficient. The third and fourth terms, proportional to $\lambda_2$ and $\lambda_3$ respectively, represent the isotropic and anisotropic parts of the force density generated by the electric field. The last term proportional to $\lambda_4$ characterizes the sensitivity of the pumping to the bending of the tissue~\cite{ramaswamy2000}.

Similarly, we can also write a constitutive equation for the electric current density $j_\alpha$: 
\begin{align}
j_\alpha\!=\!-\bar\kappa(v^{\rm c}_\alpha \!-\! v^{\rm f}_\alpha) + \Lambda_1 p_\alpha \!+ \Lambda_2 E_\alpha \!+ \Lambda_3q_{\alpha\beta}E_\beta\! +
\Lambda_4 \partial_\beta q_{\alpha\beta} \, , 
\label{eq_eletricCurrent}
\end{align}  
where $\bar\kappa$ is the coefficient that characterizes the current due to the (relative) flow of ions between cells as a consequence of a reverse electroosmotic effect~\cite{kirby2013}. The coefficient~$\Lambda_1$ characterizes the contribution of ion pumping to the electric current, while $\Lambda_2$ and $\Lambda_3$ are the isotropic and anisotropic part of the electric conductivity tensor. The coefficient $\Lambda_4$ is an active flexoelectric coefficient, indicating that a spatially nonuniform cell polarity orientation is obtained in response to an electric field\footnote{Notice that both $\lambda_4$ and $\Lambda_4$ were already introduced in~\cite{sarkar2019} but they had a vanishing contribution.}. This term plays in particular a crucial role in lumen nucleation as we explain in the following.

Finally, assuming that cells and interstitial fluid have the same mass density, and in the limit of an incompressible tissue, that we consider in the following, mass conservation can be rewritten in such a way that the total volume flux is divergence-free~\cite{ranft2012}. Considering that there is no fluid flow inside the lumen, the incompressibility yields a relation between the cell velocity and the fluid velocity inside the tissue: $v_r^{\rm f} = -\frac{\phi}{1-\phi}v^c_r$ where we have introduced the cell volume fraction $\phi$, that we assume to be constant in our model. Similarly, the charge conservation in the quasistatic limit $\partial_\alpha j_\alpha=0$ can be integrated directly in the absence of external current and yields $j_r=0$ throughout the tissue.

    \subsection{Boundary conditions}

The spheroid is surrounded by an external fluid both inside (in the lumen), and outside. This fluid exerts a hydrostatic pressure on the tissue that must be balanced by the tissue surface tension and by the total normal stress at the boundaries:
\begin{align}
-\sigma^{\rm c}_{rr}(R_1) + P^{\rm f}(R_1) &= P^{\mathrm{ext}}_1-2\gamma_1/R_1  \, , \label{eq_bc_stress} \\
-\sigma^{\rm c}_{rr}(R_2) + P^{\rm f}(R_2) &= P^{\mathrm{ext}}_2+2\gamma_2/R_2 \, , 
\end{align}
where we have introduced the inner and outer tissue surface tensions $\gamma_1$ and  $\gamma_2$. Fluid exchange between the spheroid and the outside is driven by osmotic conditions:
\begin{align}
v^{\rm f, ext}_1 \!\!- \ddt{R_1} &= \!+ \Lambda^{\rm f}_1 \left[(P^{\mathrm{ext}}_1- P^{\rm f}(R_1))-\Pi^{\mathrm{ext}}_1 \right]+J_{{\rm p},1} \, , \label{eq_bc_osmo} \\
v^{\rm f, ext}_2 \!\!- \ddt{R_2} &= \!- \Lambda^{\rm f}_2 \left[(P^{\mathrm{ext}}_2- P^{\rm f}(R_2))-\Pi^{\mathrm{ext}}_2 \right]-J_{{\rm p},2} \, .
\end{align} 
Here, $\Lambda^{\rm f}_i$ is the permeability of the interface to water flow. The fluxes $J_{{\rm p},i}=-\Lambda^{\rm f}_i (\Pi^{\mathrm{ext},0}_i-\Pi^{\mathrm{int},0}_i)$, with $\Pi^{\mathrm{ext},0}_i$ and $\Pi^{\mathrm{int},0}_i$ denoting respectively the outside and inside osmotic pressures of osmolites that can be exchanged  between external fluid and tissue, can be nonzero as a result of active pumps and transporters that maintain an osmotic pressure difference, and act effectively as water pumps. We have also defined $v^{\rm f, ext}_{1,2}$, the external flows imposed at the inner and outer boundaries. Consistently with the assumption made earlier that there is no flow inside the lumen, we consider in the following that $v^{\rm f, ext}_1= v^{\rm f, ext}_2=0$.

The normal velocity of the cells at the boundaries has to match the growth of the spheroid radii. 
An increased cell proliferation in a thin surface layer has been observed in growing spheroids~\cite{delarue2014,montel2011,delarue2013}. Thus, for the sake of generality, we allow for a thin surface layer of cells, both facing outside and to the lumen (see Fig. \ref{fig_notation}), to have a growth rate that differs from the bulk. The cell velocity boundary conditions then read:
\begin{align}
    v_r^{\rm c}(R_1) = \ddt{R_1} + v_1 \, , \\
    v_r^{\rm c}(R_2) = \ddt{R_2} - v_2 \, , \label{eq_bc_v2}
\end{align} 
where $v_i=\delta k_i  n^{\rm c}_i e/n^{\rm c}$ with $e$ the thickness of the boundary layers, $n^{\rm c}_i$ and $\delta k_i$ the cell number density and the cell growth rate in the surface layers, respectively.

\section{Lumen nucleation in a spherical cell aggregate}
    
    \subsection{Equations for the dynamics of the spheroid and its lumen}
    
In the previous section we have introduced the bulk equations that describe the properties of the tissue and can be integrated to obtain the cell velocity profile. The values of the different phenomenological parameters that we have defined can be obtained using experimental data and order-of-magnitude estimates (see Table~\ref{table_estimationParam},  App.~\ref{sec_estimations}, and Ref.~\cite{sarkar2019}). In particular, for a spheroid which typical radius is of the order of the hundred of micrometers $10^{-6} \lesssim r \lesssim 10^{-3}$~m, our estimates indicate that effects that are relevant at length scales larger than experimentally accessible ones can be neglected to obtain a simpler velocity profile (see App.~\ref{sec_derivation_radii} for details).

The bulk cell velocity profile together with the boundary conditions introduced in the previous section then allow us to obtain the dynamics of the inner~$R_1(t)$ and outer~$R_2(t)$ radii of the spheroid in the quasistatic limit. We obtain two coupled nonlinear differential equations for the spheroid radii, Eqs.~\eqref{eq_R1_adim} and~\eqref{eq_R2_adim} (see App.~\ref{sec_derivation_radii}). In these equations, six effective parameters are introduced. Two effective pressures:
\begin{align}
\begin{split}
    P^{\rm eff}_{1,2} &= \Pi_{1,2}^{\rm ext}- P_{\rm h}^{\rm c} - \frac{J_{{\rm p},1,2}}{\Lambda^{\rm f}_{1,2}} \\
    &- \frac{2}{3} \left(\zeta_0\nu_0+ \lambda_4 +\frac{\Lambda_4\lambda + (3\nu_1/2 -  2\nu_0\nu_3)\Lambda_1}{\Lambda} \right)  , \label{eq_Peff}
\end{split}
\end{align}
where $\Lambda = \Lambda_2+ 2\Lambda_3/3$ is an effective conductivity and where $\lambda  = \lambda_2+ 2\lambda_3/3$. The effective pressure $P^{\rm eff}_1$ (resp. $P^{\rm eff}_2$) can be seen as a modification of the homeostatic pressure~$P^{\rm c}_{\rm h}$ by the external osmotic pressure $\Pi_1^{\rm ext}$ (resp. $\Pi_2^{\rm ext}$), the pumping flux $J_{{\rm p},1}$ (resp. $J_{{\rm p},2}$) and electric and active contributions. In particular, if all other quantities are kept constant, we observe that a positive effective pressure $P^{\rm eff}_1>0$ indicates a stress on the spheroid inner boundary leading to a shrinkage of the tissue, similar to the increase in cell apoptosis due to a pressure larger than the homeostatic pressure in simpler settings~\cite{ranft2010,ranft2012}. Three apparent tension parameters are also introduced:
\begin{align}
    \gamma^{\rm app}_{1,2} \!\! = \! \gamma_{1,2} \! \mp \! 4\nu_3\Lambda_4/\Lambda \, , \, \gamma^{\rm app}_0 \!\!= \! \left( 3\nu_1/2 \!- \! 2(2+\nu_0)\nu_3 \right) (\Lambda_4/\Lambda) \! \, . \label{eq_gammaEff}
\end{align}
The apparent tension $\gamma^{\rm app}_1$ (resp. $\gamma^{\rm app}_2$) of the inner (resp. outer) surface is a modification of the tissue surface tension $\gamma_1$ (resp. $\gamma_2$) stemming from the flexoelectric term proportional to $\Lambda_4$ -- which generates an electric field as a result of curvature -- and the field-induced anisotropic stress characterized by $\nu_3$. This modification can in principle lead to a negative apparent surface tension for the inner or outer part of the spheroid. A negative inner apparent surface tension ($\gamma^{\rm app}_1<0$) will be shown in the following to enhance lumen formation and allow for spontaneous nucleation. The apparent tension parameter $\gamma^{\rm app}_0$ is also due to flexoelectricity. It enters the dynamical equations of the inner and outer radii of the spheroid in a similar way as the apparent surface tensions (see Eqs.~\eqref{eq_R1_adim} and~\eqref{eq_R2_adim} in App.~\ref{sec_derivation_radii}). Finally, an effective pumping coefficient is introduced:
\begin{align}
    \lambda^{\rm eff}=\lambda_1- \Lambda_1 \lambda/\Lambda \, , \label{eq_lambdaEff}
\end{align}
which is a combination of the active pumping term $\lambda_1$ and of an electric contribution $\Lambda_1 \lambda/\Lambda$ to the pumping due to electroosmotic effects. This term comes as a prefactor of the spheroid thickness $R_2-R_1$, indicating that the whole tissue acts as a pump.

   \subsection{Lumen nucleation: a competition between pumping and electric effects}
   \label{sec_nucleation}
   
We now discuss lumen formation in spherical cell aggregates. Assuming that the radius of the lumen is small at the early stage of its formation, we can cast the lumen early dynamics into the form of a nucleation problem. In particular, we highlight in the following the crucial role in lumen formation of pumping and of the active flexoelectrity which contributes to the apparent surface tension $\gamma^{\rm app}_1$. In the small lumen limit $R_1 \ll R_2$, the equations describing the radii dynamics partially decouple and we obtain the following dimensionless equations:
\begin{align}
    \ddroit{r_1}{\hat t}  &=  \delta_1 - \frac{2 \hat \gamma_1}{r_1}-\frac{a}{r_2}+ \frac{b-\delta_2 \, r_2+ \hat\lambda\, r_2^2/4}{r_2(3+ \chi r_2)} + \frac{3}{4} \hat\lambda \, r_2  \, ,\label{eq_r1Thick} \\
    \ddroit{r_2}{\hat t} &= \frac{b-\delta_2 \, r_2+ \hat\lambda \, r_2^2/4}{3+ \chi r_2} \, ,\label{eq_r2Thick} 
\end{align}
where we have introduced dimensionless radii: $r_i(\hat t)= R_i(t)/ R_0 $ with $R_0=\Lambda^{\rm f}_1 \bar \eta$ and a dimensionless time $\hat t=t/\tau_0$ with $\tau_0=\bar\eta/|P^{\rm eff}_1|$. We have also introduced the dimensionless parameters:
\begin{align}
    & \hat \gamma_{0,1,2} \!=\! \frac{\gamma^{\rm app}_{0,1,2}}{\bar \eta \Lambda^{\rm f}_1 |P^{\rm eff}_1|} \, , \quad \hat \lambda  \!=\!   \frac{ \lambda^{\rm eff} \bar\eta \Lambda^{\rm f}_1 }{|P^{\rm eff}_1|} \, ,  \, \quad \delta_{1,2}  \!=\! \frac{P^{\rm eff}_{1,2}}{ |P^{\rm eff}_1|} \, , 
\label{eq_dimless_parameters}
\end{align}
and $\chi\!=\!\Lambda^{\rm f}_1/\Lambda^{\rm f}_2$, $a\!=\!3\hat v_2-2\hat\gamma_0$, $b \!=\! 3\hat v_2-2(\hat\gamma_2+\hat\gamma_0)$, $\hat v_2 \!=\! v_2/(\Lambda^{\rm f}_1 |P^{\rm eff}_1|)$. Note that the inner surface growth velocity $v_1$ does not contribute to lumen nucleation. Indeed, lumen growth is mainly fed by inward fluid flow, while an increased cell division at the inner surface comes as a lower order effect.

To allow for an analogy with nucleation of a droplet in a fluid, the equation for the dynamics of $r_1$ can be rewritten as:
\begin{align}
    \dd r_1/ \dd \hat t = f(r_2)-2\hat\gamma_1/r_1 \, , \label{eq_nucleation}
\end{align}
with $f(r_2)=\delta_1-a/r_2+(b-\delta_2 r_2+ \hat\lambda r_2^2/4)/(r_2(3+\chi r_2))+3\hat\lambda r_2/4$. In this form, we can make an analogy with the nucleation of a droplet in a fluid: lumen nucleation is driven by a competition between the bulk contribution (first term in Eq.~\eqref{eq_nucleation}) and the surface term (second term in Eq.~\eqref{eq_nucleation}). To continue the analogy with droplet nucleation, we also introduce a lumen critical radius:
\begin{align}
    r_1^c=2\hat\gamma_1/f(r_2) \, , \label{eq_criticalRadius}
\end{align}
which is the radius above which a lumen starts growing. Notice that in this nonequilibrium, active system, the lumen critical radius depends on the value of the outer radius $r_2$, and therefore on time. Moreover, the (dimensionless) apparent surface tension $\hat\gamma_1$ can be negative as the result of active flexo\-electricity, and in this case a lumen can open  spontaneously starting even from a vanishingly small radius.

\begin{figure}[t]
	\centering 
	\null \hfill
	{\includegraphics[width=0.48\linewidth]{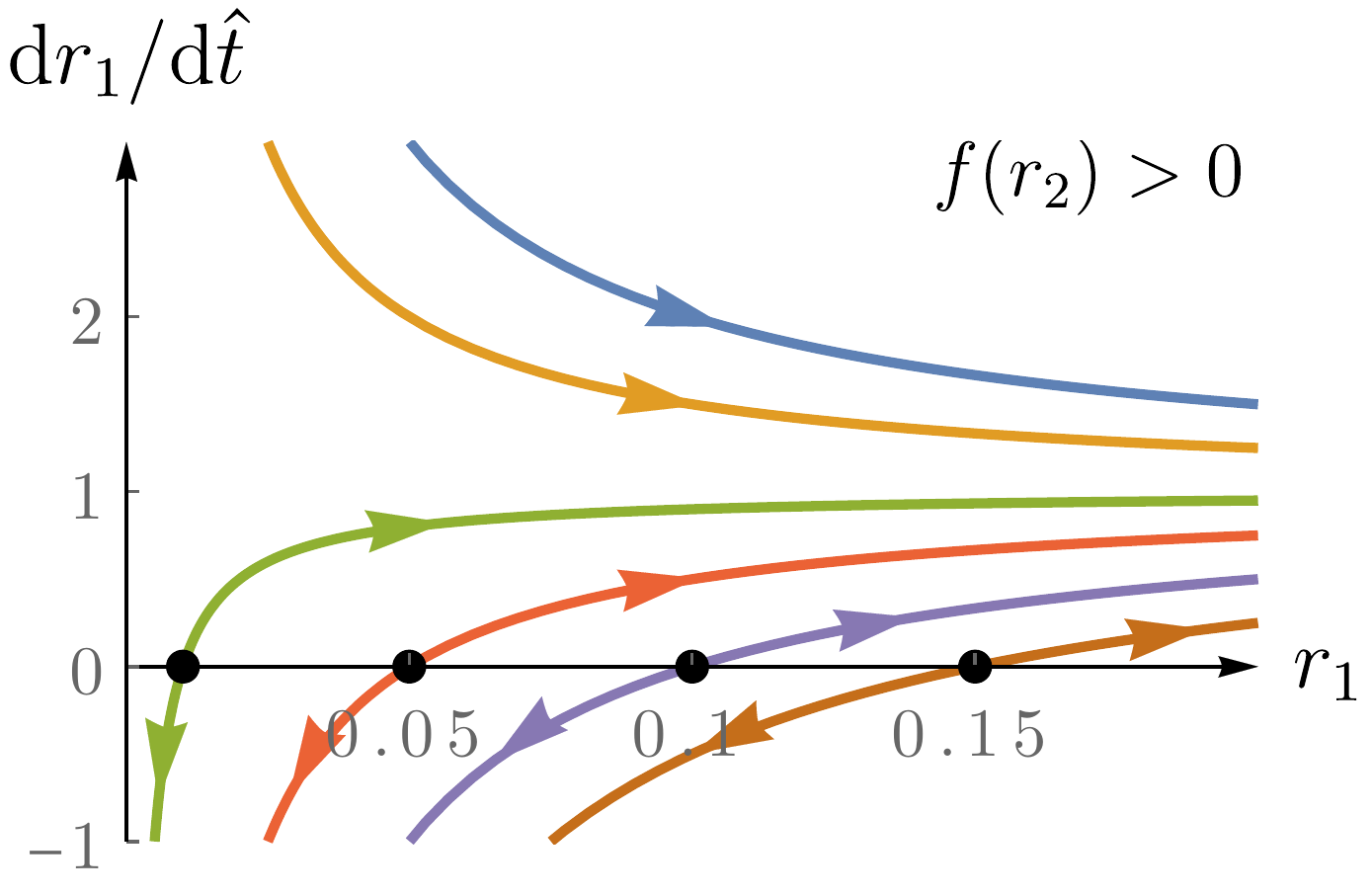}}
    \hfill
    {\includegraphics[width=0.48\linewidth]{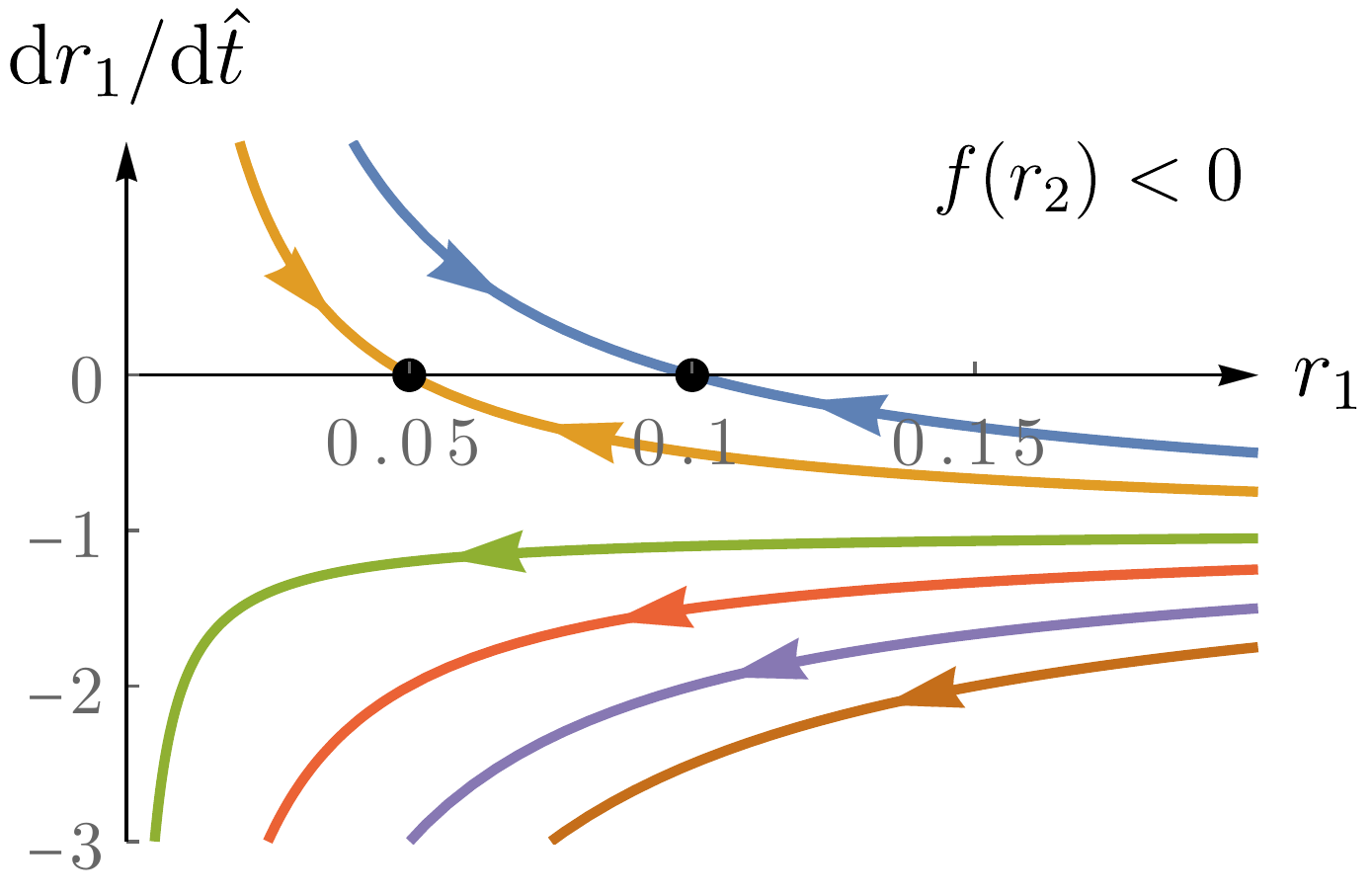}}
    \hfill \null
    \caption{Phase portrait $(r_1,\dd r_1/ \dd \hat t)$ for different values of the apparent surface tension~$\hat\gamma_1$. \textbf{Left:} The lumen state is favored ($f(r_2)>0$), and lumens with radius larger than the critical radius $r_1^c$ (black dots) grow, while lumens with smaller radius shrink, as indicated by the arrows. However, as the apparent surface tension is lowered (from lower brown to upper blue curve), the surface cost for creating a lumen is lowered, and eventually vanishes: a lumen spontaneously forms (uppermost yellow and blue curves without fixed point). \textbf{Right:} The lumen state is not favored ($f(r_2)<0$), and a lumen with a large radius is always prohibited. As long as the apparent surface tension is positive (lower curves), a lumen is always unstable and shrinks (no fixed point). However, as the apparent surface tension is lowered (from lower brown to upper blue curve) and becomes negative, an attractive fixed point can be found at a finite radius (black dots) which means that a lumen of finite size spontaneously grows. Plots were obtained by setting $\hat\gamma_1=\{0.15, 0.1, 0.05, 0.01, -0.05, -0.1\}$ (from lower brown to upper blue curve), and $f(r_2)=1$ (left panel) or $f(r_2)=-1$ (right panel).}
    \label{fig_functional}
\end{figure}

Let us now be more specific and consider the case where $r_2$ is kept fixed, and the volume contribution $f(r_2)$ is positive. The phase portrait of the system for this case is plotted on the left panel of Fig.~\ref{fig_functional}. Let us first emphasize that the condition $f(r_2)>0$ can always be satisfied if the effective pumping term $\hat \lambda$ is positive (inward pumping) and if $r_2$ is sufficiently large: a signature that the aggregate acts collectively for providing fluid to the lumen. According to Eq.~\eqref{eq_lambdaEff}, a positive effective pumping can be achieved with a positive active fluid pumping coefficient $\lambda_1$, indicating an inward pumping, or with a negative ion pumping coefficient $\Lambda_1$, indicating an inward pumping of ions, or with both. Because of this pumping, the larger the spheroid (that is, $r_2$), the more cells contribute to the inward flow, and therefore the smaller the critical radius $r_1^c$. This effect is so dramatic that for an arbitrarily small early lumen $r_1(t=0)=\varepsilon$, one can always find a large enough spheroid such that $\varepsilon>r_1^c$ and thus such that the lumen starts growing. Notice also that a positive effective pressure ($P^{\rm eff}_1>0$, which implies $\delta_1>0$), which is a modification of the homeostatic pressure, indicates an unfavorable environment for cells in the center and is therefore favorable for lumen formation.

Moreover, when the apparent surface tension is positive ($\gamma^{\rm app}_1>0$, which implies $\hat\gamma_1>0$), the usual droplet nucleation picture is preserved in the sense that the system must first perform work against the apparent surface tension (nucleation barrier) to effectively form a growing lumen. However, the value of $\gamma^{\rm app}_1= \gamma_1 - 4\nu_3\Lambda_4/\Lambda$ is a competition between the tissue surface tension $\gamma_1$ and the active flexoelectric contribution $4\nu_3\Lambda_4/\Lambda$, which effectively lowers the tissue surface tension. As $\hat\gamma_1$ is decreased --~which can for instance be achieved by reducing the tissue surface tension at the boundary with the lumen~$\gamma_1$, or by increasing the tissue flexoelectric coefficient~$\Lambda_4$~-- the cost for nucleation is  lowered (black dots on the left panel of Fig.~\ref{fig_functional}) until it eventually vanishes for $\gamma^{\rm app}_1\leq0$ (which implies $\hat\gamma_1\leq0$). As a result, the nucleation barrier vanishes as the apparent surface tension changes sign.

The case $f(r_2)<0$ is also illuminating: in this case, the usual droplet nucleation picture would indicate that the droplet phase is not stable since it does not lower the system energy. The phase portrait for this scenario is plotted on the right panel of Fig.~\ref{fig_functional}: when the apparent surface tension is positive ($\gamma^{\rm app}_1>0$, which implies $\hat\gamma_1>0$), there is no fixed point and the system is always driven to the lumenless state, as one would expect in the droplet nucleation picture. However, as soon as the apparent surface tension becomes negative, a new attractive fixed point exists at a finite radius, meaning that a small lumen forms spontaneously. In this case, the negative surface tension term drives lumen formation even if a lumen is disfavored by the volume term.

Whether the volume term favors a lumen state or not, we find that a lumen spontaneously nucleates whenever its apparent surface tension becomes negative ($\gamma^{\rm app}_1 < 0$), that is when
\begin{align}
    4\nu_3\Lambda_4/\Lambda>\gamma_1 \, . \label{eq_spontaneousNucleation}
\end{align}
This happens when the flexoelectric effects (proportional to $\Lambda_4$) overcome the usual tissue surface tension. 

For fixed values of $r_2$, we have seen that lumen formation can be cast in the form of a nucleation problem. In fact, one can even introduce the (dimensionless) volume $V_1=4\pi r_1^3/3$ and the function $\Psi (r_1) = 4\pi r_1^2 \hat\gamma_1  - (4/3) \pi r_1^3 f(r_2)$ such that one has $\dd V_1/ \dd \hat t = -\partial_{r_1}\Psi (r_1)$ and the lumen radius is obtained by minimizing $\Psi$ at fixed $r_2$. The final picture is however more subtle because the outer radius $r_2$ is time dependent and evolves according to Eq.~\eqref{eq_r2Thick}. The curl of $(\dd r_1/ \dd \hat t,\dd r_2/ \dd \hat t)$ does not vanish and the dynamics given by equations~\eqref{eq_r1Thick} and~\eqref{eq_r2Thick} thus does not result from the gradient of an effective potential. Indeed, the nonpotential nature of the dynamics allows for thickness oscillations of the spheroid, as we discuss in the next section.

\section{Spheroid and lumen dynamics at long time}
\label{sec_spheroidFate}

    \subsection{Spheroid long-time states}
    
\begin{figure}[t]
	\centering 
	\null \hfill
    \subfigure[\label{fig_LgrowthSgrowth}]
    {\includegraphics[width=0.24\linewidth]{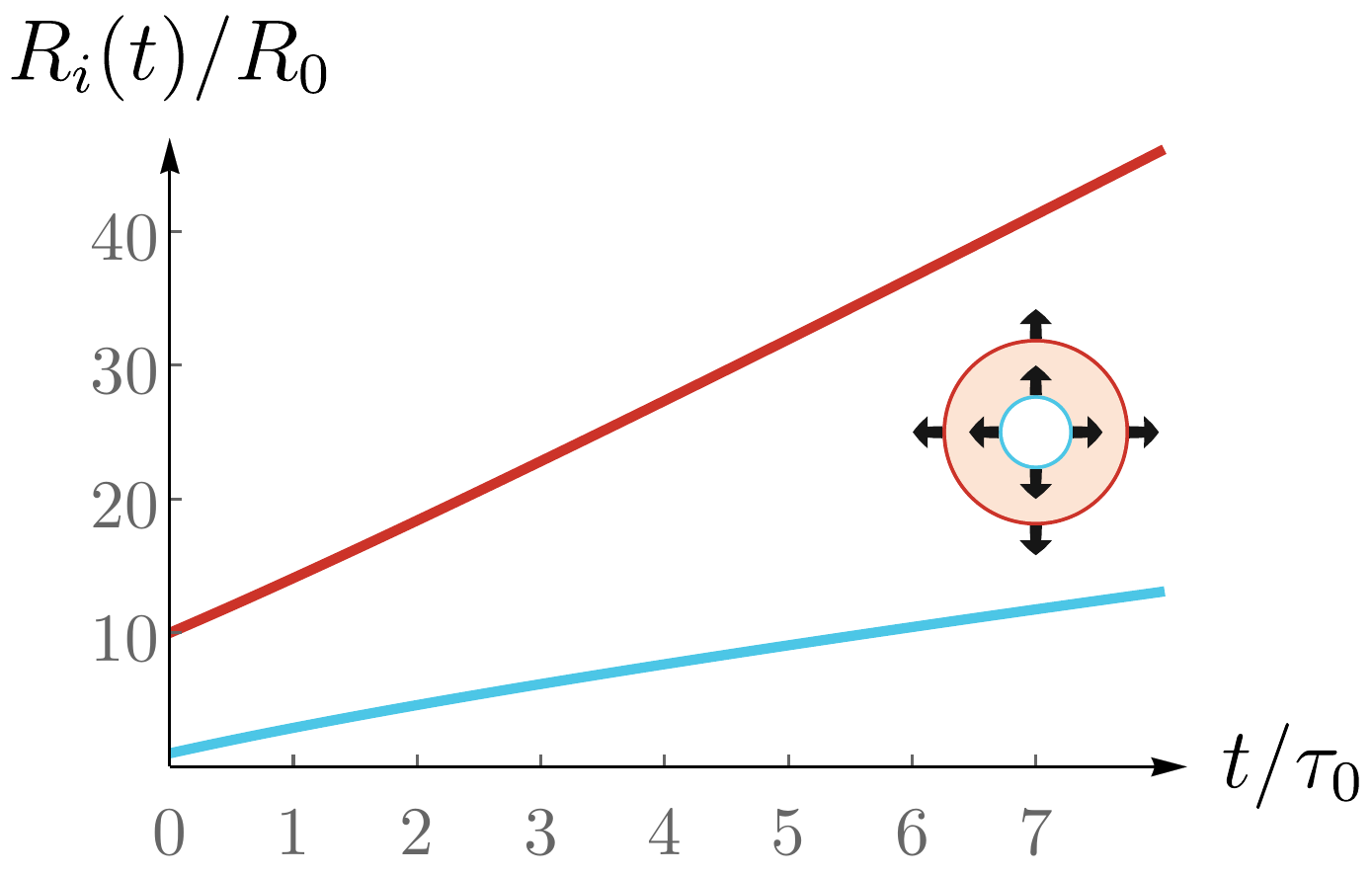}}
    \hfill
	\subfigure[\label{fig_LsteadySgrowth}]    {\includegraphics[width=0.24\linewidth]{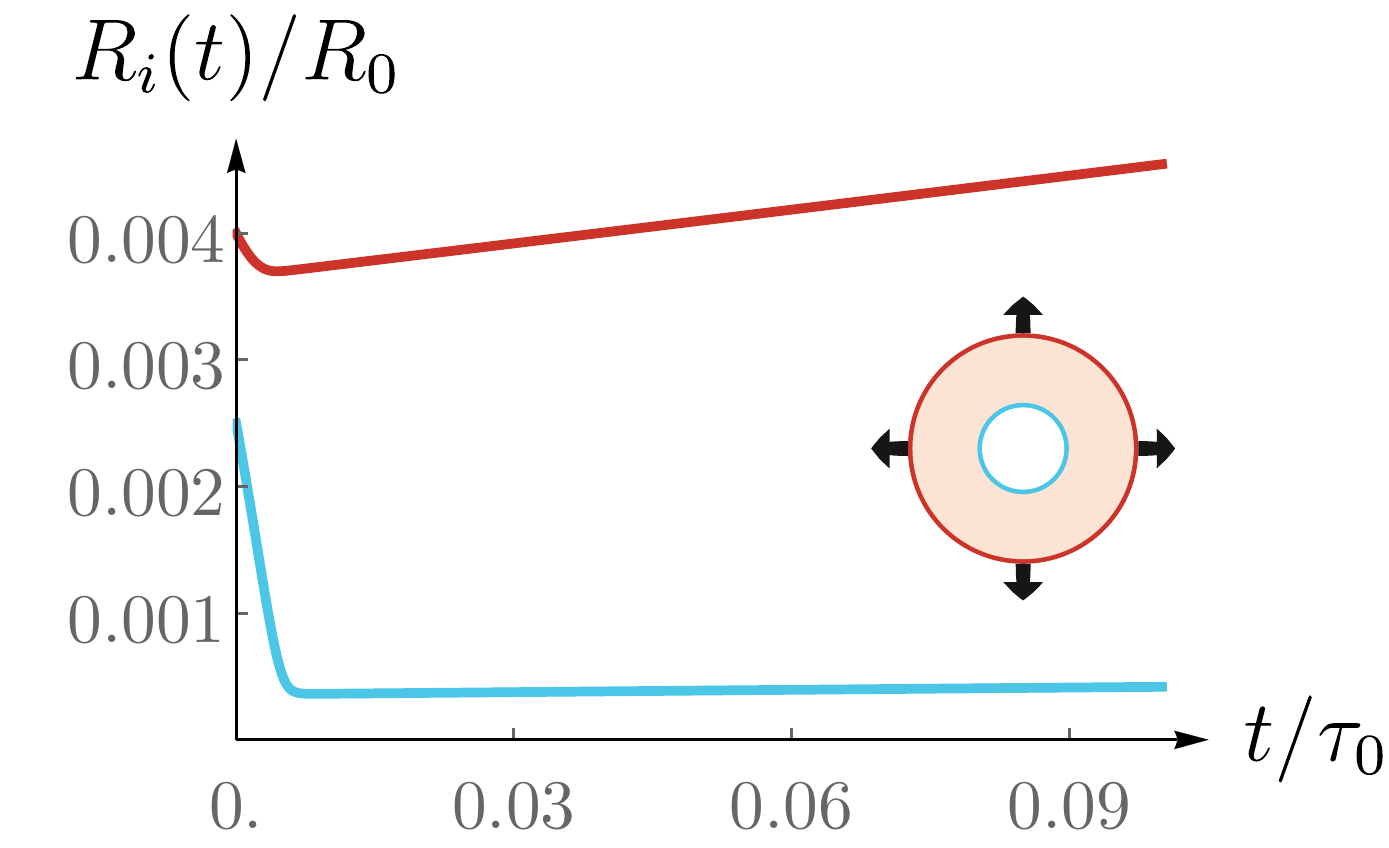}}
    \hfill
    \subfigure[\label{fig_ssWithLumen}]
    {\includegraphics[width=0.24\linewidth]{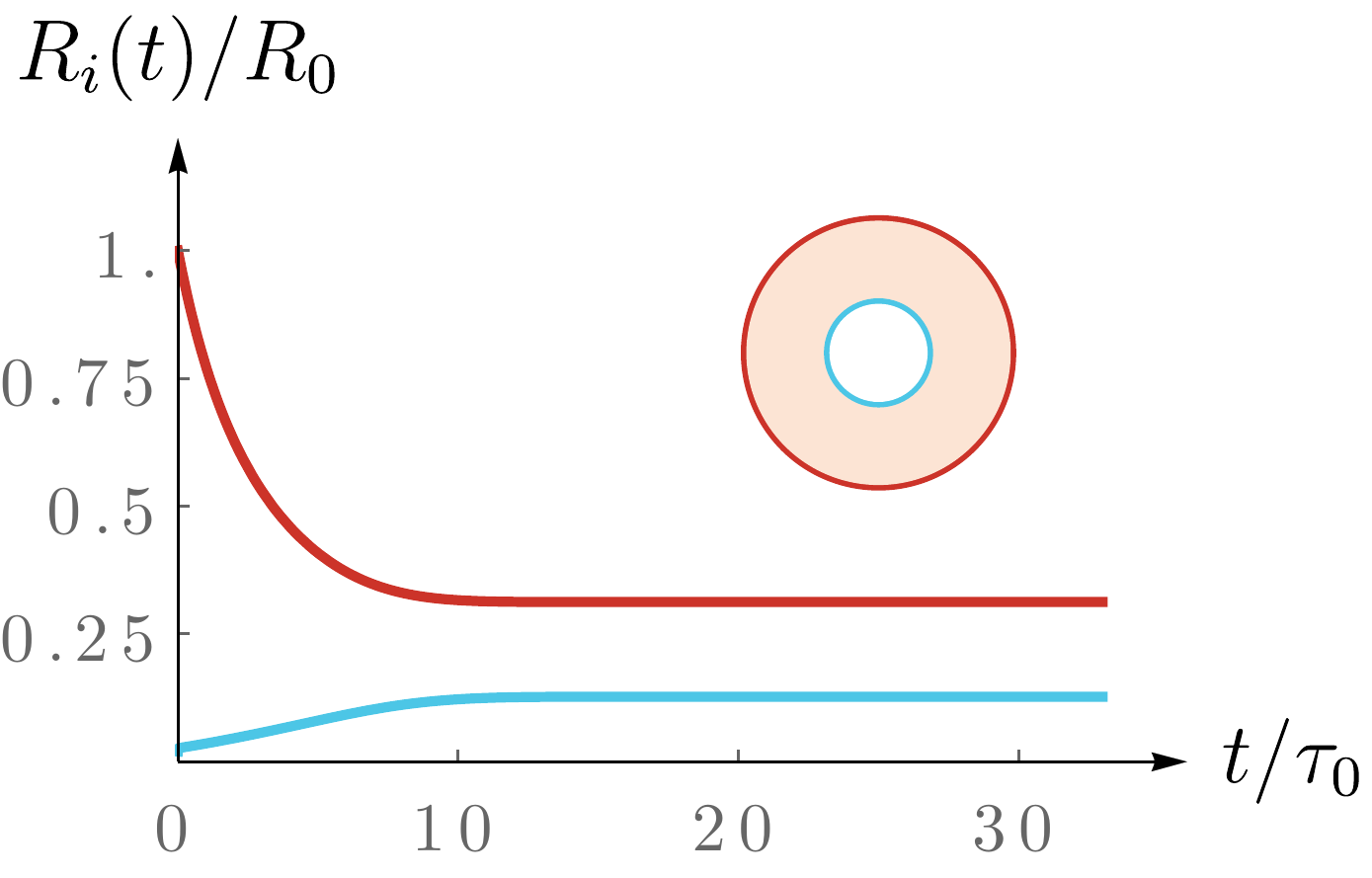}}
    \hfill
    \subfigure[\label{fig_decay_to_ssMonolayer}]
    {\includegraphics[width=0.24\linewidth]{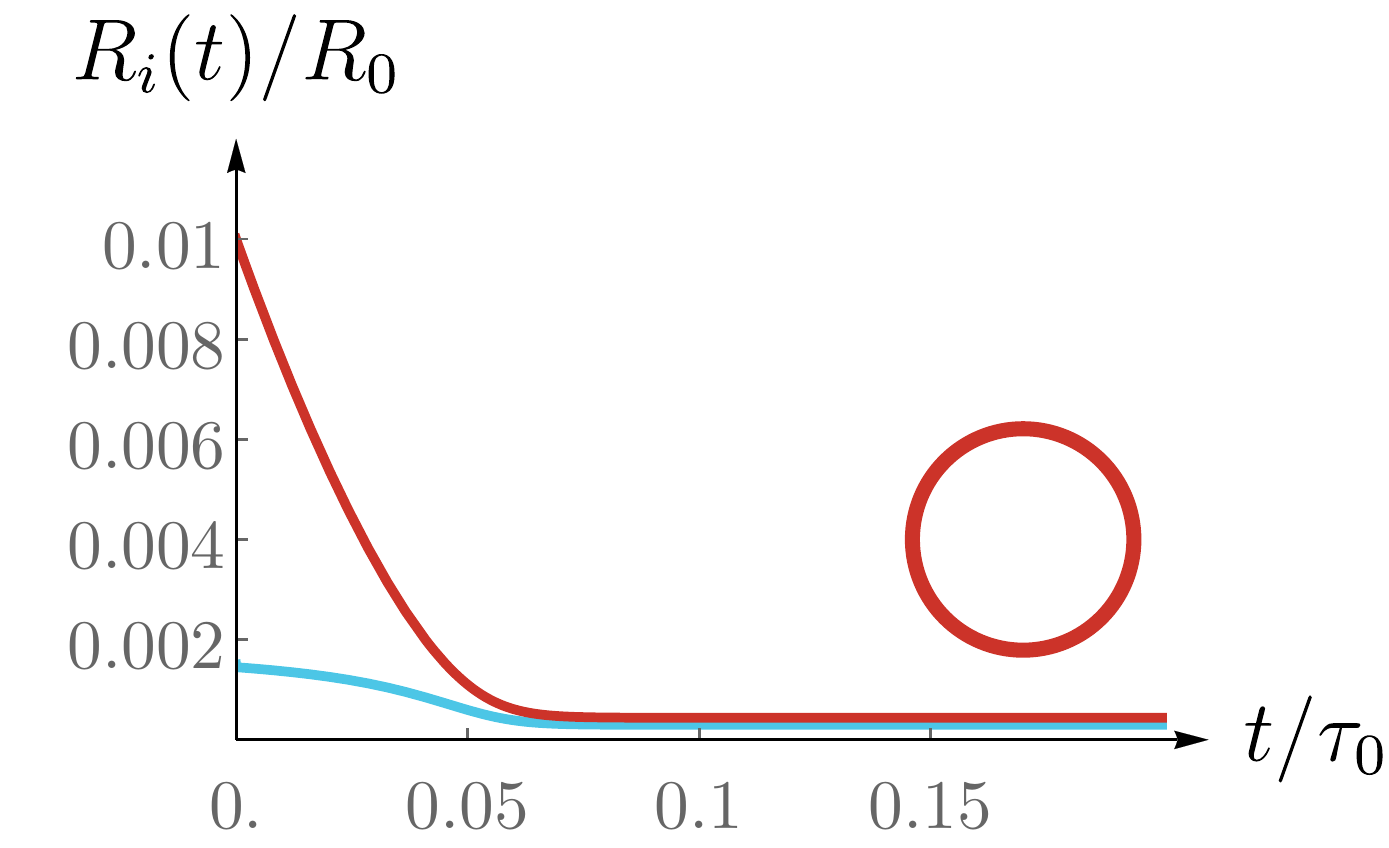}}
    \hfill \null
    
	\null \hfill
	\subfigure[\label{fig_oscillations}]
    {\includegraphics[width=0.24\linewidth]{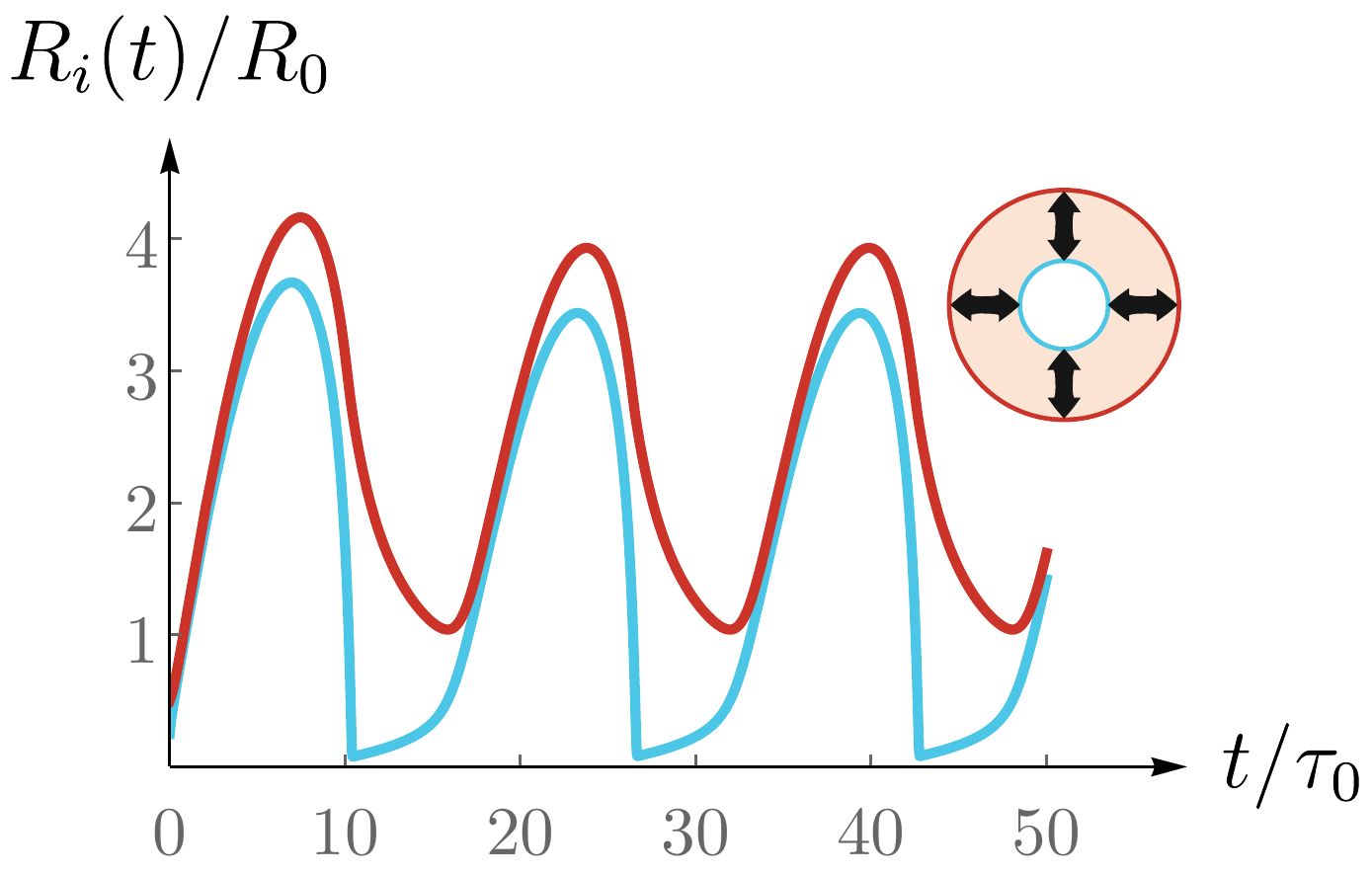}}
    \hfill
    \subfigure[\label{fig_lumenClosure}]
    {\includegraphics[width=0.24\linewidth]{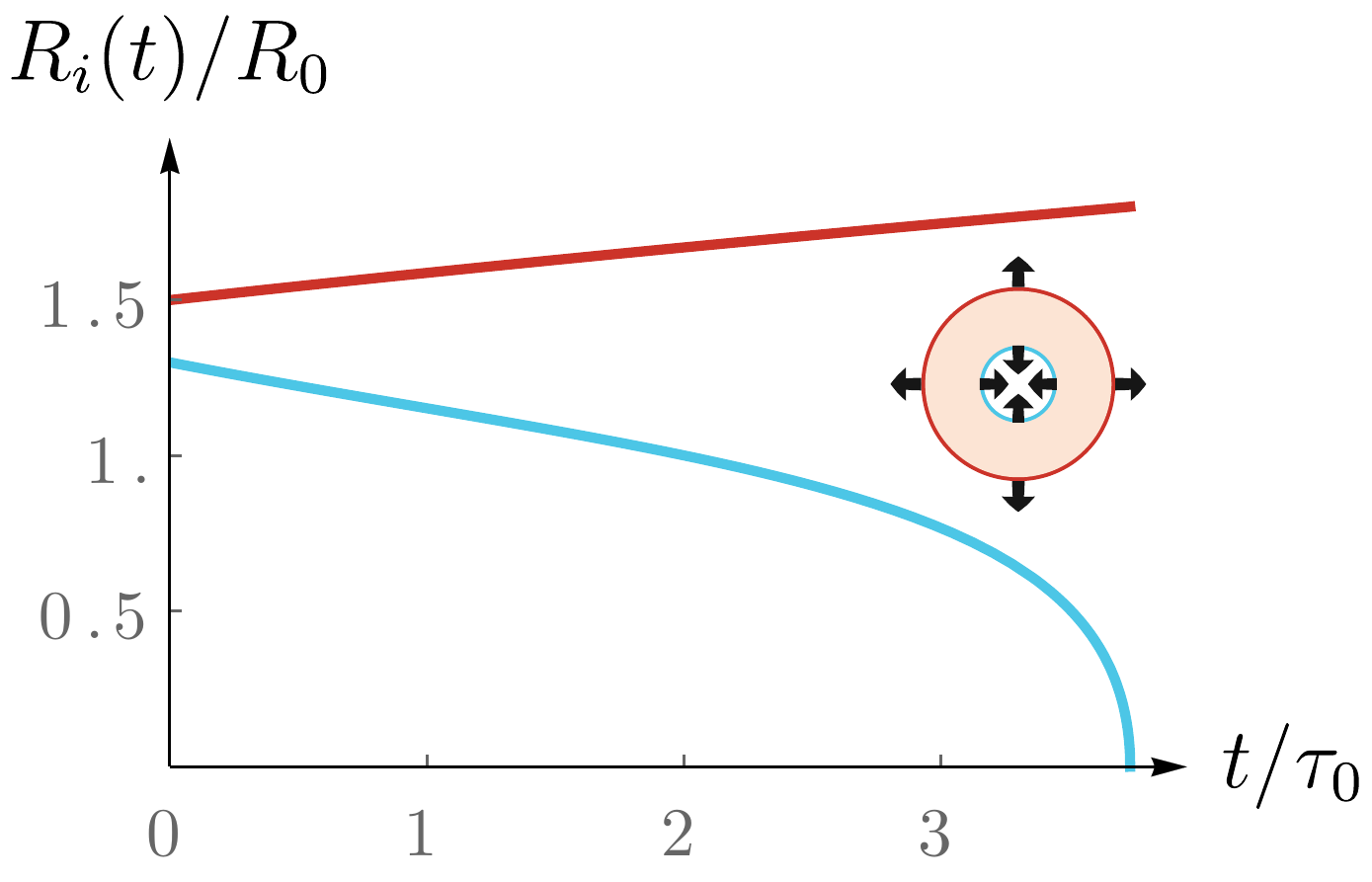}}
    \hfill
    \subfigure[\label{fig_growth_to_monolayer}]
    {\includegraphics[width=0.24\linewidth]{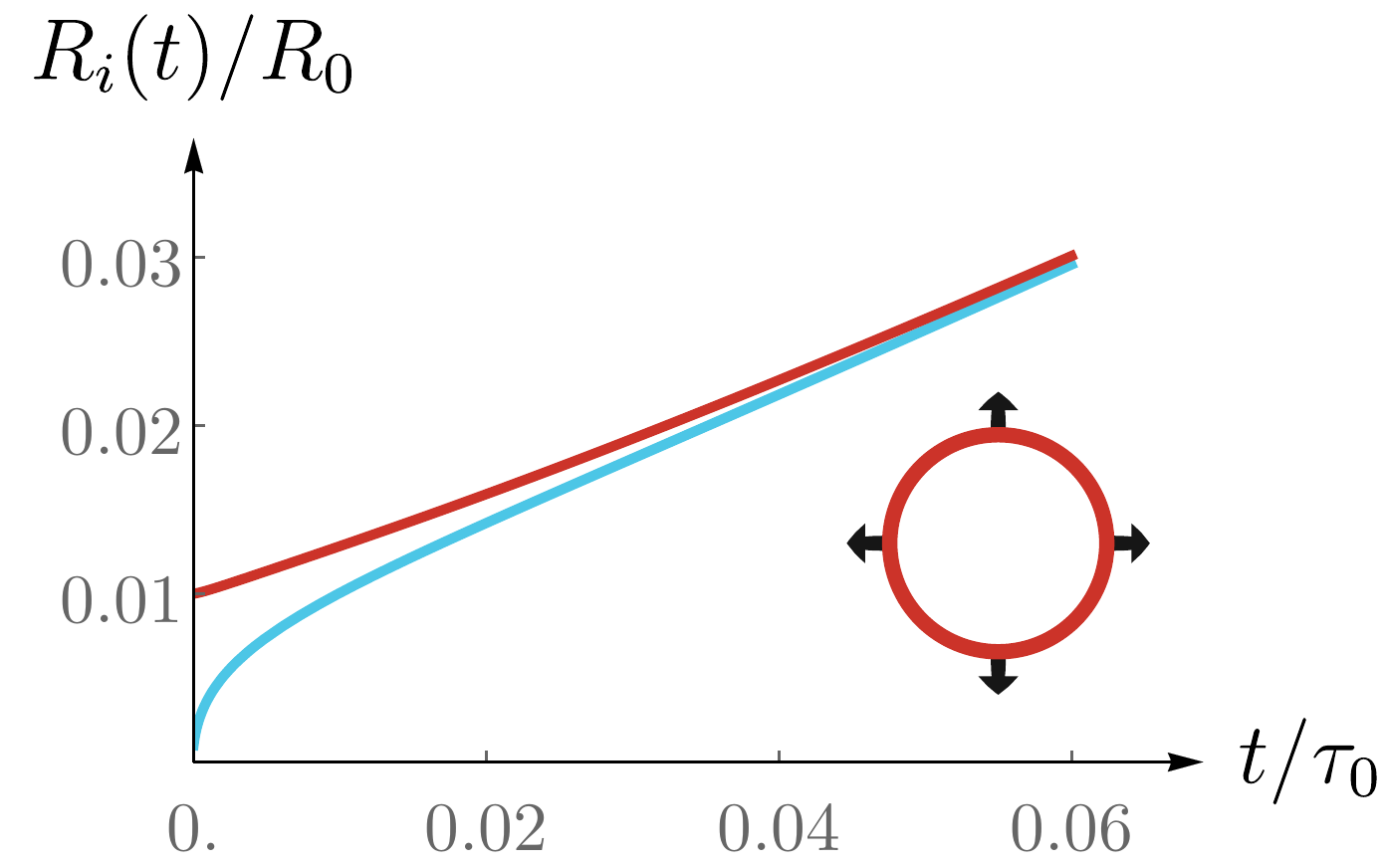}}
    \hfill
    \subfigure[\label{fig_decay_to_monolayer_singular}]
    {\includegraphics[width=0.24\linewidth]{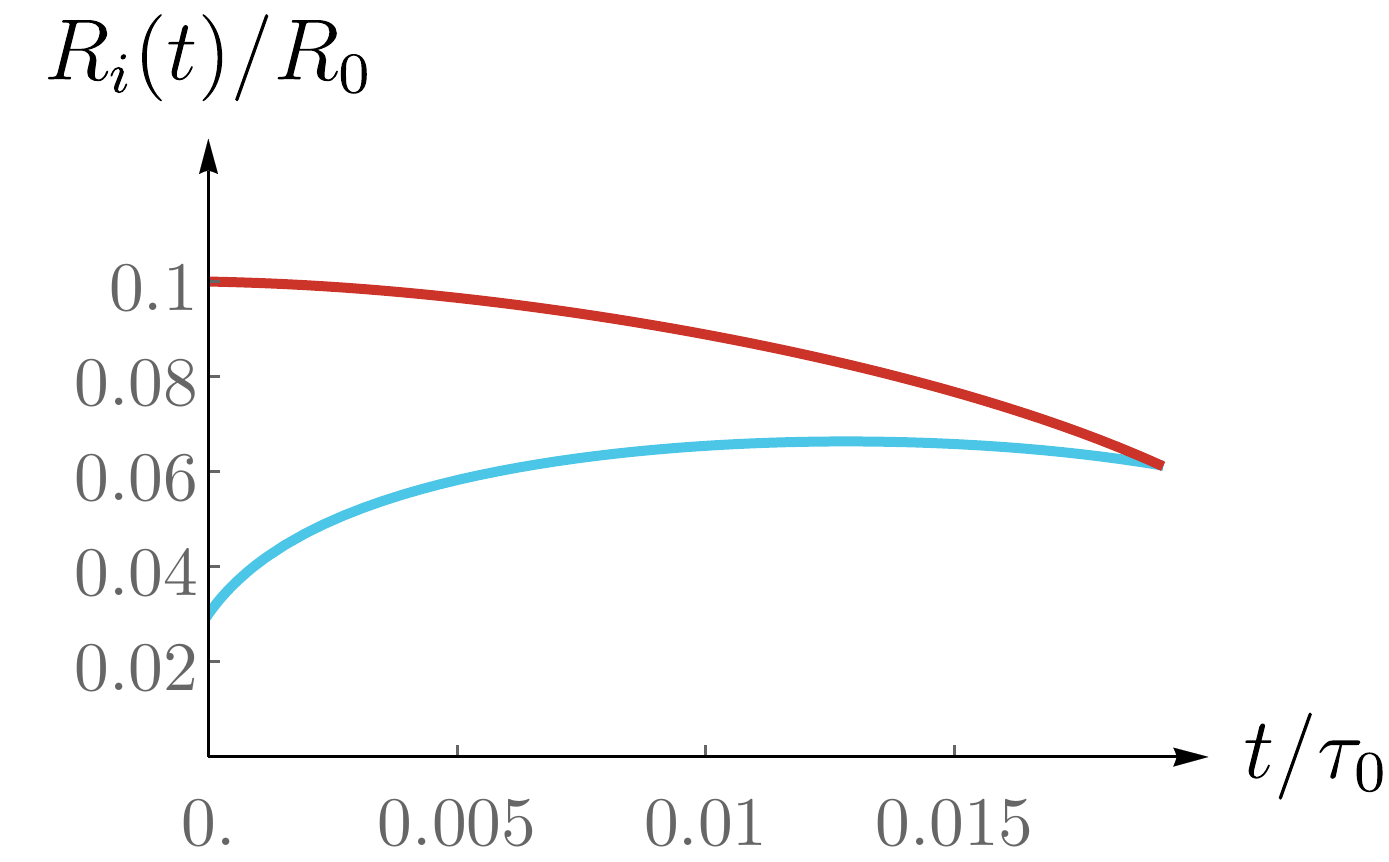}}
    \hfill \null
    \caption{Examples of different dynamics of cell spheroids. Inner ($R_1$, lower blue curve) and outer ($R_2$, upper red curve) spheroid radii as a function of time for different values of the parameters, illustrating the long-time states of the system. The plots are obtained by solving numerically equations~\eqref{eq_R1_adim} and \eqref{eq_R2_adim}. (a) Spheroid and lumen growth. (b) Spheroid growth and lumen steady state. (c) Spheroid and lumen steady state. (d) Spheroid and lumen decay leading to a steady-state quasi-monolayer. (e) Thickness oscillations of the spheroid. (f) Lumen closure and spheroid growth. (g) Spheroid and lumen growth leading to a growing monolayer. (h) Spheroid decay is faster than that of the lumen, leading to a collapse of the whole structure. We have rescaled time and length units using $\tau_0=\bar\eta/|P^{\rm eff}_1|$ and $R_0=\bar \eta\Lambda^{\rm f}_1$. Refer to Table~\ref{table_parameters_plots} in App.~\ref{sec_estimations} for the values of the parameters used to obtain these plots.}
    \label{fig_overview}
\end{figure}

After lumen nucleation, the spheroid and its lumen follow a dynamics that depends on the parameters of the model. The spheroid and its lumen keep evolving according to Eqs.~\eqref{eq_R1_adim} and \eqref{eq_R2_adim}, and the diversity of the long-time scenarii is illustrated in Fig.~\ref{fig_overview}, where the time evolution of the spheroid radius $R_2$ and the lumen radius $R_1$ are plotted for parameter values illustrating the different regimes. 

In the case where lumen nucleation is favored, which can be achieved for instance with a negative apparent inner surface tension~$\hat\gamma$, corresponding to conditions where the flexoelectric coupling overcomes the inner tissue surface tension, several long-time fates are possible for the cell aggregate. The spheroid and its lumen can grow (see Fig.~\ref{fig_LgrowthSgrowth}): this is achieved in this example because a positive inner effective pressure $\delta_1$ favors lumen growth and a negative outer effective pressure $\delta_2$ favors the outer radius growth. Switching the inner effective pressure~$\delta_1$ to negative values, the lumen can reach a steady state while the spheroid grows (Fig.~\ref{fig_LsteadySgrowth}). Alternatively, both lumen and spheroid can reach steady states (Figs.~\ref{fig_ssWithLumen} and \ref{fig_decay_to_ssMonolayer}).  This is for instance seen in the case of an outward effective pumping of fluid ($\hat\lambda<0$) which limits the spheroid growth. Under conditions where lumen formation is favored by flexoelectricity while the pumping is directed outwards, the spheroid can even undergo thickness oscillations if the time scales associated with these mechanisms are sufficiently different (Fig.~\ref{fig_oscillations}).

When the lumen is not favored, it shrinks and eventually vanishes (Fig.~\ref{fig_lumenClosure}). In this example the coefficients describing active pumping and flexoelectricy are zero, and the apparent inner surface tension is positive. Figure~\ref{fig_growth_to_monolayer} shows an example of a growing spheroid with growing lumen where the lumen radius approaches the outer radius. This could correspond to the formation of a single layer spheroid. Finally,  Fig.~\ref{fig_decay_to_monolayer_singular} provides an example of  spheroid where the outer layer shrinks faster than the lumen, leading to the disappearance of the spheroid when $R_1=R_2$. This behavior results from a positive outer effective pressure, $\delta_2>0$. Similar behavior can also occur when the lumen grows faster than the outer layer, for example if both inner and outer surface growth velocities~$\hat v_{1,2}$ are negative.

    \subsection{State diagram of the spheroid dynamics}
    
\begin{figure}[t]
	\centering 
	\null \hfill
    {\includegraphics[width=0.95\linewidth]{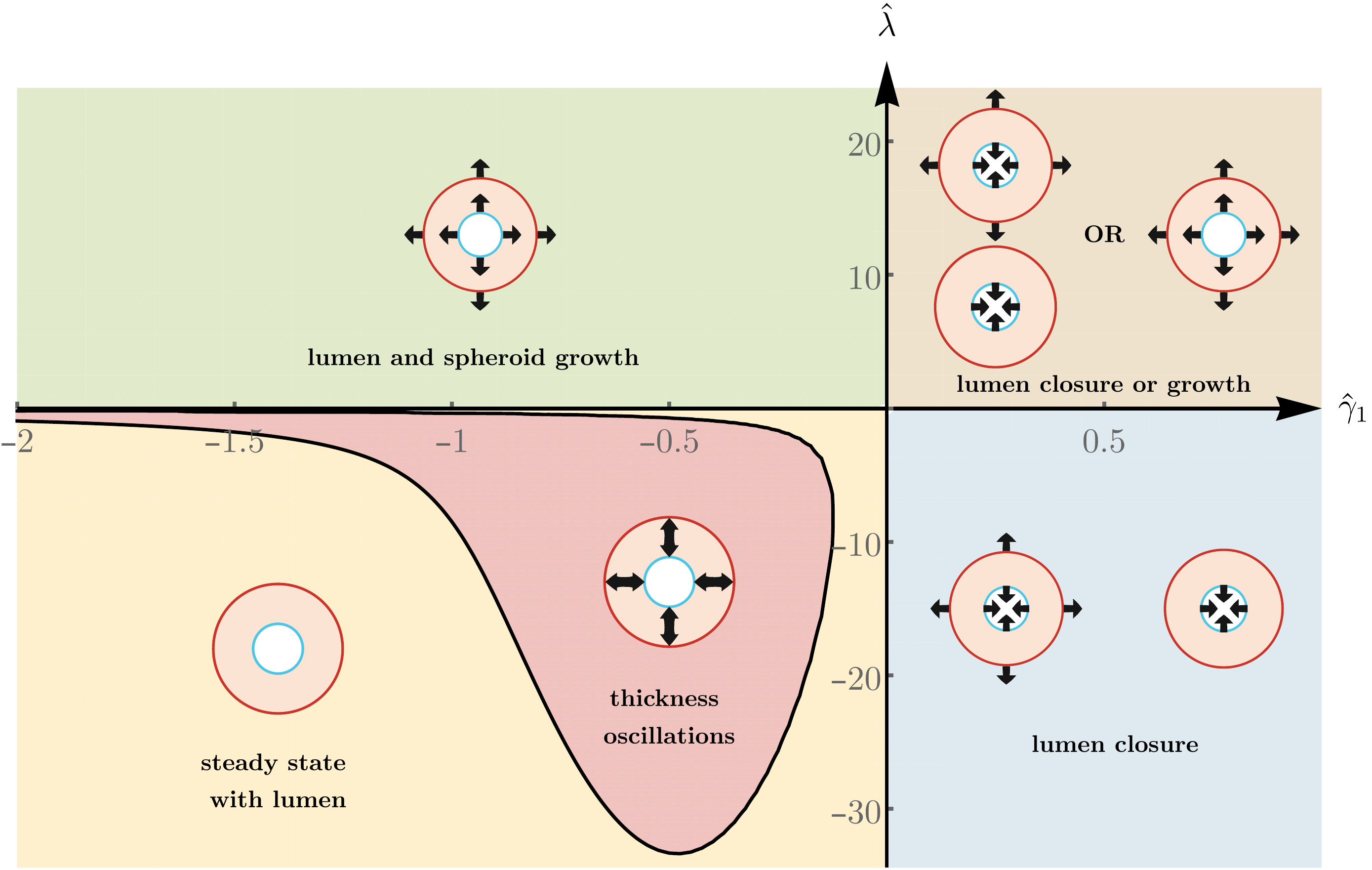}}
    \hfill \null
    \caption{Typical state diagram of spheroid dynamics as a function of the dimensionless effective pumping strength $\hat\lambda$ and apparent inner surface tension $\hat\gamma_1$. Several regions with different types of behaviors are displayed. Upper-left region (green): the spheroid and its lumen grow. Upper-right region (brown): the spheroid lumen may either shrink and eventually close or grow depending on the initial radii. Lower-right region (blue): the lumen always shrink and disappear. Lower-left region (yellow): the spheroid and its lumen reach a steady state with finite radii. Lower-left region (red): the spheroid undergoes thickness oscillations. Solid black lines (including the $x$ and $y$ axes) indicate boundaries between regions. For details see text. Parameter values are: $\delta_1\!=\!1$, $\delta_2\!=\!-1$, $\hat v_1\!=\!\hat v_2\!=\!0.1$, $\chi\!=\!1$, $\hat \gamma_0=0$ and $\hat \gamma_1\!+\!\hat \gamma_2\!=\!0.1$.
    }
    \label{fig_stateDiagram}
\end{figure}

We now discuss a typical state diagram to illustrate parameter regions where different behaviors displayed in Fig.~\ref{fig_overview} can be found. In the state diagram shown in Fig.~\ref{fig_stateDiagram}, we vary the dimensionless apparent inner surface tension $\hat \gamma_1$ and the dimensionless effective pumping coefficient $\hat \lambda$, while the other parameters are held fixed. We consider the case where the effective pressure $P^{\rm eff}_1$ is fixed. The fixed parameters are chosen as follows. We consider a negative $\delta_2$ and positive $\delta_1$, promoting growth of the outer radius $R_2$ and of the lumen $R_1$, respectively (see Eqs.~\eqref{eq_R1_adim} and \eqref{eq_R2_adim}). The values of the surface layer growth velocities $\hat v_{1,2}$ are chosen positive to match experimental results~\cite{delarue2014,montel2011,delarue2013}. Their precise values play a minor role in the lumen dynamics, which is mostly controlled by the fluid influx and outflux across the surface of the cellular aggregate. Inner and outer surface permeabilities are chosen to be equal such that $\chi=1$. The apparent tension $\hat\gamma_0$ is set to 0 as it plays a marginal role in lumen nucleation (see Eqs.~\eqref{eq_r1Thick} and \eqref{eq_r2Thick}). Finally we choose the sum of the inner and outer tissue surface tensions, such that $\hat\gamma_1+\hat\gamma_2=0.1$. These dimensionless values correspond to physical quantities given in Table \ref{table_estimationParam}.

For negative apparent inner surface tensions $\hat \gamma_1<0$, which corresponds to the left half plane of the state diagram, a lumen always forms, see Eq. \eqref{eq_r1Thick}. If in addition the effective pumping $\hat \lambda$ is positive (inward pumping), lumen and spheroid grow and there is no steady state. If instead, $\hat \lambda$ is negative (outward pumping), steady state spheroids can occur but also oscillations of the spheroid in certain parameter ranges. Onset of spheroid oscillations corresponds to the appearance of a stable limit cycle in the dynamics and is delimited in the diagram by the solid black curve between the yellow and red regions in the lower left quadrant. As discussed in App.~\ref{sec_oscillations}, the system undergoes a Hopf bifurcation when going from the steady-state region to the oscillation region. For positive apparent inner surface tensions $\hat \gamma_1>0$, lumen tends to shrink with one exception. If in addition $\hat \lambda$ is negative, lumen always shrink while the spheroid can be stationary or growing. Finally, if both $\hat \gamma_1$ and $\hat \lambda$ are positive, they have opposite effects on lumen size, and the outcome depends on the initial radii $R_1$ and $R_2$ and on parameters values. If the initial radius of the lumen is smaller than the critical radius $R_1^c$ (which is a function of $R_2$), the lumen shrinks, while if this initial radius is larger than this critical value, the spheroid and the lumen grows. For sufficiently large outer radius $R_2$, a lumen always grows.

The values of~$\hat \gamma_1$ and~$\hat \lambda$ depend on the physical coefficients describing fluid and ion pumping, flexoelectricity and surface tension. In the case where the effective pressure~$P^{\rm eff}_1$ is kept constant, the apparent inner surface tension~$\hat \gamma_1$ depends linearly on the tissue inner surface tension~$\gamma_1$ and on the flexo\-electric coupling $\Lambda_4$ (see Eq.~\eqref{eq_gammaEff}). Similarly, the effective pumping coefficient $\hat \lambda$ depends linearly on the fluid pumping coefficient $\lambda_1$ and on the ion pumping coefficient $\Lambda_1$ (see Eq.~\eqref{eq_lambdaEff}). For instance, an increase in the flexoelectric coupling~$\Lambda_4$ can result in a sign change from positive to negative of the apparent inner surface tension~$\hat\gamma_1$ and thus promote lumen nucleation. Similarly, changing the direction of fluid or ion pumping from inwards to outwards, which corresponds to a sign change of the coefficients~$\lambda_1$ or~$\Lambda_1$, respectively, can result in a sign change of the effective pumping coefficient~$\hat\lambda$. In the case of positive apparent inner surface tension, this sign change leads to lumen suppression. Conversely in the case of negative apparent inner surface tension, this leads to the existence of a steady state with finite radii or to thickness oscillations.

Notice that spheroid oscillations have been observed both \textit{in vitro}~\cite{ruiz-herrero2017} and \textit{in vivo}~\cite{futterer2003} in the case of monolayers, but to the best of our knowledge this phenomenon has not been observed in thick spheroids. For monolayered spheroids, oscillations are usually explained by a cycle of growth of the spheroid that builds a stress on the shell, which finally burst open and therefore shrinks due to the outward fluid flow. This is followed by a healing of the hole and the whole process repeats~\cite{futterer2003,ruiz-herrero2017}. For the thick spheroids we consider here, such a bursting process cannot take place, but interestingly oscillatory regimes can still be predicted. 

In the state diagram shown in Fig.~\ref{fig_stateDiagram}, we find most of the behaviors shown as examples in Fig.~\ref{fig_overview}. However the behavior of a steady-state lumen radius while the spheroid grows indefinitely (Fig.~\ref{fig_LsteadySgrowth}) requires a different choice of parameters. Indeed, a necessary condition to find this specific behavior is $\delta_1<0$ (together with $\hat\lambda=0$, $\delta_2>0$ and $\hat\gamma_1<0$).

%% file: conclusion.tex
\section{Conclusion}

Our theoretical work has shown that the formation of a fluid-filled lumen in cell assemblies is governed by nucleation equations that resemble those describing the nucleation of a droplet in a fluid. The nucleation of a lumen however depends on additional effects that are fundamentally active: tissue response to mechanical stress, tissue fluid and ion pumping and tissue active flexoelectricity. 
In the present context, flexoelectricity describe the ability of a polar tissue to generate an electric current when the polarity axis is splayed. This effect could also be observed for nonpolar cells, provided they have an axis of anisotropy. One might expect that such effects are small compared to those generated by uniform polarity, but our findings here show that they can be significant in the nucleation process of lumen. One possible mechanism for generating a flexoelectric current results from the wedge shape of cells in a splayed tissue, leading to a difference in ion pumping on basal and apical sides and thus in electric current generation.

In addition to exhibiting the role of the coupling between mechanical, hydraulic and electrical mechanisms in lumen formation, we have also used our model to explore the role of this coupling in the long-time dynamics of the aggregate and its lumen. In particular, tissue active flexoelectricity, associated with the mechano-electric response of the tissue, is revealed to play a crucial role in early lumen formation as it generates a bulk term in Eq.~\eqref{eq_gammaEff} that acts as a surface tension. Our order-of-magnitude estimations (see Table~\ref{table_estimationParam}) indicate that this effect is significant, and the flexoelectric contribution could overcome the tissue surface tension, leading to a negative apparent surface tension. Such a negative apparent inner surface tension guarantees the nucleation of a lumen. Note that the negative contribution to the apparent surface tension stems from a bulk stress proportional to $1/R$ rather than from a genuine surface tension. Therefore negative apparent surface tension should not lead to surface shape instabilities.

Similarly, we observe that the fluid pumping is also influenced by electric effects: the effective pumping term that is defined in Eq.~\eqref{eq_lambdaEff} contains a contribution that stems directly from active cell pumping, but also an additional one which can be understood as an electroosmotic contribution to the active pumping. Electroosmotic flows are generated when an electric field is applied to a fluid close to a charged surface~\cite{marbach2019} and recent studies suggest that these electroosmotic flows could be for instance dominant in the corneal fluid transport~\cite{sanchez2002,fischbarg2017}. In our analysis, an inward effective pumping ensures lumen nucleation if the spheroid is large enough. We have moreover shown that active tissue flexoelectricity and pumping can work in concert, in which case lumen formation is facilitated (or prohibited if both effects tend to close the lumen). However, if fluid pumping is directed outwards while flexoelectric effects tend to open a lumen, steady states of lumen and spheroid are observed. This antagonism between flexoelectric and pumping effects can also give rise to electrohydraulic oscillations, which are radically different from the oscillations observed for spherical cell monolayers that rely on a burst and healing mechanism~\cite{futterer2003,ruiz-herrero2017,chan2019}

Because our framework is based on symmetry considerations and does not rely on specific cell-based mechanisms, we expect our results to give a robust qualitative picture of lumen formation in cell assemblies. Below the cell scale, other approaches based on similar principles become relevant \cite{dasgupta2018}. Lumen nucleation in the mouse embryo has for instance been observed to rely on the nucleation and coarsening of multiple micrometer-sized lumens~\cite{dumortier2019}. Once these micrometer-sized lumens have fused, our approach should capture its further evolution. Lumen formation in a cell aggregate can also lead to the formation of a monolayer spheroid~\cite{andrew2010,sigurbjornsdottir2014,datta2011}. Such monolayers arise in our analysis in certain regimes, see Figs.~\ref{fig_decay_to_ssMonolayer} and \ref{fig_growth_to_monolayer} for instance.

Finally, the role of mechanical stress on tissue morphogenesis, the importance of electric effects and fluid transport in cells and tissues are well-known facts. However, we have highlighted in this paper the interplay between these effects and how they need to be combined to understand lumen formation in cell assemblies. The potential importance of such an interplay is suggested for instance in a recent work on the zebrafish fin regeneration~\cite{daane2018}, which shows the importance of potassium channels in growth phenotypes. This suggests that electric effects may couple to growth process and size control of tissues~\cite{blackiston2009,cervera2018,levin2018}. Such observations therefore open the door to studies of tissue morphogenesis where hydraulic, electric and mechanical effects are brought together.

%% file: acknow.tex
\subsection*{Acknowledgments}

JP dedicates his contribution to Prof. Peter S. Pershan (Harvard University) on his 85th birthday, remembering his mentoring in soft and condensed matter physics. We thank Keisuke Ishihara and Arghyadip Mukherjee for insightful discussions. CD thanks Szabolcs Horv\'at for his MaTeX package that was used to produce the figures.

%% file: appendices.tex
\section{Derivation of the dynamics equations for the inner and outer radii}
\label{sec_derivation_radii}

Using the constitutive equations introduced in the main text, we can rewrite the force balance equation \eqref{eq_forceBalance_spherical} as a differential equation on the cell velocity only. The equation we obtain is the starting point of our study and reads:
\begin{align}
    \eta_1 v(r)\left(\frac{1}{r^2}+\frac{1}{L_0^2}+\frac{1}{L_1 r} \right) +\eta_2 v'(r) \left(  \frac{1}{L_2}- \frac{1}{r}  \right)-\eta_3 v''(r) = \tilde\lambda_1 + \frac{2\tilde\zeta}{r} + \frac{2 \tilde\gamma}{r^2}  \, , \label{eq_vcells}
\end{align}
where here and in the following we drop the superscript $\rm c$ for the cell velocity. We have introduced the following effective lengths:
\begin{align}
    L_0^{-1} &= \sqrt{\frac{\alpha \kappa^{\rm eff}}{\eta_1 (1-\phi)}} \, ,  \quad 
    L_1^{-1} = \frac{2}{\eta_1 (1-\phi)}  \left( 
    \frac{(\zeta_1 \nu_1 - 2\nu_3)\bar \kappa}{\Lambda}
     + \zeta_1 \nu_2 - 2\nu_4
    \right)  \, ,\\
    L_2^{-1}  &= \frac{1}{\eta_2 (1-\phi)}  \left[ (1-2\zeta_1/3) \left( \nu_1 \frac{\bar \kappa}{\Lambda}+\nu_2 \right) + \frac{4}{3}(1-\nu_0) \left( \nu_3 \frac{\bar \kappa}{\Lambda}+\nu_4 \right) \right] \, ,
\end{align}
we have also defined effective viscosities:
\begin{align}
    \eta_1 &= (2+8\zeta_1/3)\bar \eta+4\eta(2+\nu_0)/3 \, ,\\
    \eta_2 &= (2-10\zeta_1/3)\bar \eta+4\eta(2+\nu_0)/3 \, , \\
    \eta_3 &= (1-2\zeta_1/3)\bar \eta+4\eta(1-\nu_0)/3 \, ,
\end{align}
and effective parameters:
\begin{align}
    &\lambda  = \lambda_2+ \frac{2}{3}\lambda_3 \, , \quad \Lambda = \Lambda_2+ \frac{2}{3}\Lambda_3 \, , \quad  \kappa^{\rm eff} = \kappa -\bar\kappa \frac{\lambda}{\Lambda} \, , \\
    &\tilde\gamma  = \left(\frac{4}{3}(2+\nu)\nu_3-(1+4\zeta_1/3)\nu_1 \right) \frac{\Lambda_4}{\Lambda} \, , \quad    \tilde\zeta  = \zeta_0  + \tilde \lambda_4 + (2\nu_3-\zeta_1\nu_1)\frac{\Lambda_1}{\Lambda}   \, , \\
    &\alpha  = 1-\frac{2}{3}\nu_0\zeta_1 \, , \quad  \tilde \lambda_1  = \alpha  \left( \lambda_1 - \Lambda_1 \frac{\lambda}{\Lambda} \right) \, ,  \quad \tilde \lambda_4 = \alpha \left( \lambda_4 - \Lambda_4 \frac{\lambda}{\Lambda} \right) \, . 
\end{align}
Let us now discuss the values of the effective lengths $L_0$, $L_1$ and $L_2$ that appear naturally in Eq.~\eqref{eq_vcells}. We focus in this paper on the dynamics of a spherical cell aggregate whose typical radius is of the order of the hundred of micrometers: $10^{-6} \lesssim r \lesssim 10^{-3}$~m. Using experimental data and order-of-magnitude estimations (see Ref.~\cite{sarkar2019} and App.~\ref{sec_estimations}), we are able to give estimations for the effective lengths $L_0$, $L_1$ and $L_2$ appearing in Eq.~\eqref{eq_vcells}. The permeation length $L_0$ is of the order of the millimeter and may become relevant for larger spheroids, while $L_1$ is of the order of the centimeter and $L_2$ of the order of the decimeter. In the following, we take the limit $r\ll L_i$ and therefore neglect the contribution to the dynamics of the terms involving the effective lengths $L_i$. In this limit, Eq.~\eqref{eq_vcells} becomes
\begin{align}
    -\eta_1 v(r)/r^2 +\eta_2 v'(r)/r + \eta_3 v''(r) + \tilde\lambda_1 + \frac{2\tilde\zeta}{r} + \frac{2 \tilde\gamma}{r^2} =0 \, . \label{eq_v_simple}
\end{align}
This equation can be solved using a power-law ansatz: 
\begin{align}
    v(r)=A_1 r^{\beta_1}+A_2 r^{\beta_2}+ k_0 + k_1 r + k_2 r^2 \, , \label{eq_powerLawAnsatz}
\end{align}
where $A_1$ and $A_2$ are integration constants to be determined by the boundary conditions at $r=R_1$ and $r=R_2$ (see Eqs.~\eqref{eq_bc_stress}-\eqref{eq_bc_v2}), while the coefficients $k_i$ are obtained by finding a particular solution to Eq.~\eqref{eq_v_simple} and read:
\begin{align}
    k_0 = \frac{2\tilde\gamma}{\eta_1} , \, \quad k_1 = \frac{2\tilde\zeta}{\eta_1-\eta_2} \, , \quad k_2 = \frac{\tilde\lambda_1}{\eta_1-2(\eta_2+\eta_3)} \, , 
\end{align}
and where the exponents $\beta_{1,2}$ are obtained by solving the homogeneous equation and read:
\begin{align}
    \beta_{1,2} &= \frac{1}{2}\left(1-\frac{\eta_2}{\eta_3}\mp \sqrt{1+\left(\frac{\eta_2}{\eta_3}\right)^2+4\frac{\eta_1}{\eta_3}-2\frac{\eta_2}{\eta_3}} \, \right)  \, .
\end{align}
In order to keep the analysis simpler, we have considered in the main text the limit where the active stress does not depend on the cell pressure, that is $\zeta_1\to0$. We have moreover considered the case where the tissue shear viscosity is small compared to the tissue bulk viscosity ($\eta\ll\bar\eta$), which is justified by experimental values of these parameters, indicating $\eta\simeq 10^4$~Pa$\cdot$s while $\bar\eta\simeq 10^9$~Pa$\cdot$s. In this limit, we have in particular $\beta_1=-2$ and $\beta_2=1$. Using the boundary conditions, we finally obtain the following dimensionless equations for the radii dynamics:
\begin{align}
    \begin{split}
    &\ddroit{r_1}{\hat t} + \frac{3(\hat v_1+\ddthat{r_1})r_1^2}{r_2^3-r_1^3} + \frac{3(\hat v_2-\ddthat{r_2}) r_2^2}{r_2^3-r_1^3} = \delta_1  \\
    &- \frac{2}{r_1}\left(\hat \gamma_1 -\hat \gamma_0 \frac{r_1 (r_1+r_2)}{r_1^2+r_2^2+r_1 r_2} \right) + \hat \lambda (r_2-r_1) \frac{r_1^2+2r_1 r_2+3 r_2^2}{4(r_1^2+r_2^2+r_1 r_2)} , \label{eq_R1_adim}
\end{split} \\
\begin{split}
    &-\chi \ddroit{r_2}{\hat t} +  \frac{3(\hat v_1+\ddthat{r_1})r_1^2}{r_2^3-r_1^3}+\frac{3(\hat v_2-\ddthat{r_2}) r_2^2}{r_2^3-r_1^3} = \delta_2 \\
    & + \frac{2}{r_2}\left(\hat \gamma_2+\hat\gamma_0 \frac{r_2 (r_1+r_2)}{r_1^2+r_2^2+r_1 r_2} \right) - \hat\lambda (r_2-r_1) \frac{3 r_1^2+2r_1 r_2 + r_2^2}{4(r_1^2+r_2^2+r_1 r_2)} , \label{eq_R2_adim}
\end{split}
\end{align}
where we have introduced dimensionless radii: $r_1(\hat t)= R_1(t)/ R_0 $ and $r_2(\hat t)= R_2(t)/ R_0 $ with $R_0=\Lambda^{\rm f}_1 \bar \eta$ and a dimensionless time $\hat t=t/\tau_0$ with $\tau_0=\bar\eta/|P^{\rm eff}_1|$. The dimensionless parameters have been defined in Eq.~\eqref{eq_dimless_parameters} (and $\hat v_1 = v_1/\Lambda^{\rm f}_1 |P^{\rm eff}_1|$).


  \section{Bifurcations and thickness oscillations}
  \label{sec_oscillations}
        
\begin{figure*}[t]
	\centering 
	\null \hfill
	\subfigure[\label{fig_eigenvalues}]   
	{\includegraphics[width=0.5\linewidth]{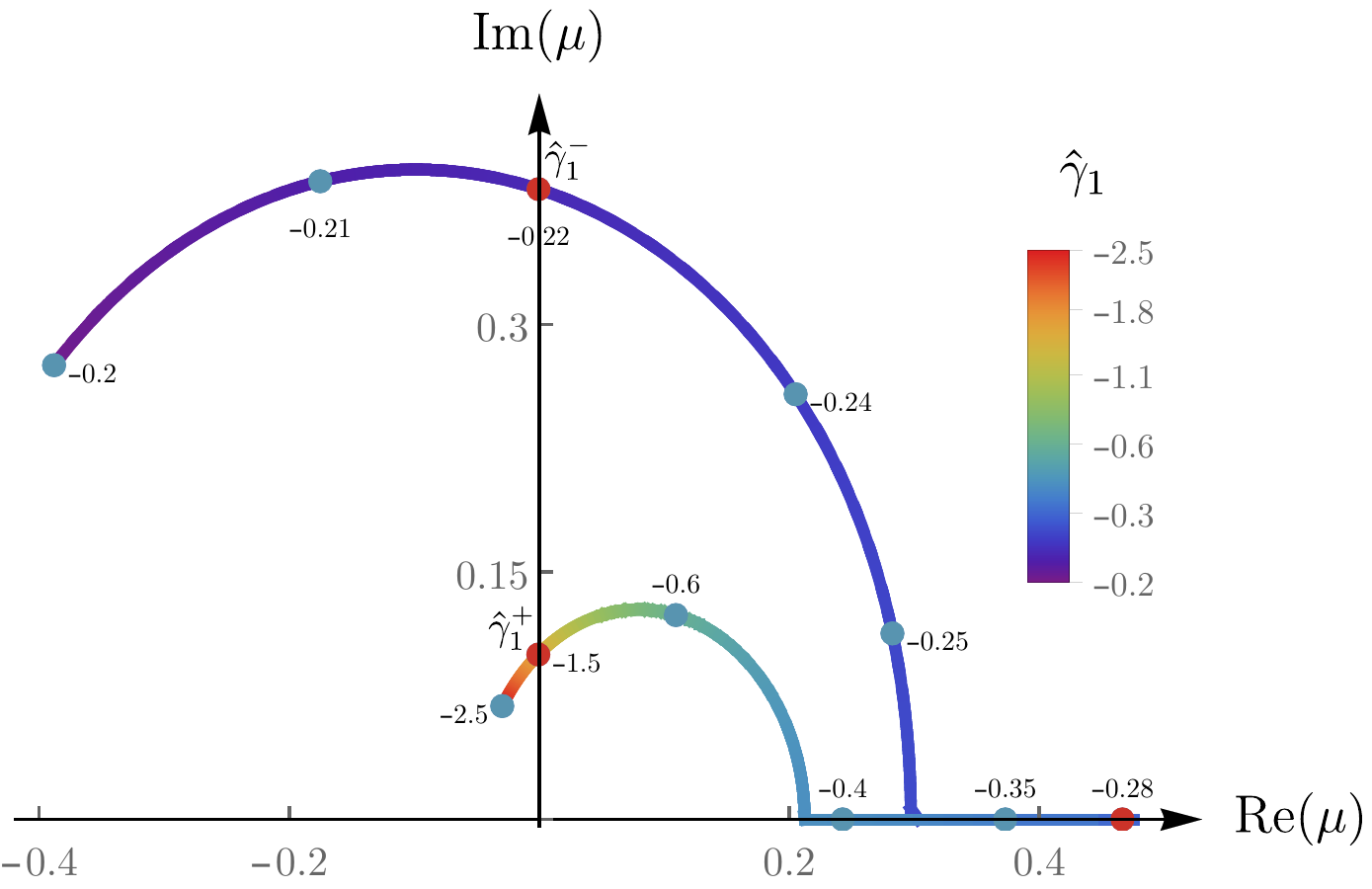}}
	\hfill \null
	
	\null \hfill
	\subfigure[\label{fig_osci1}]    {\includegraphics[width=0.32\linewidth]{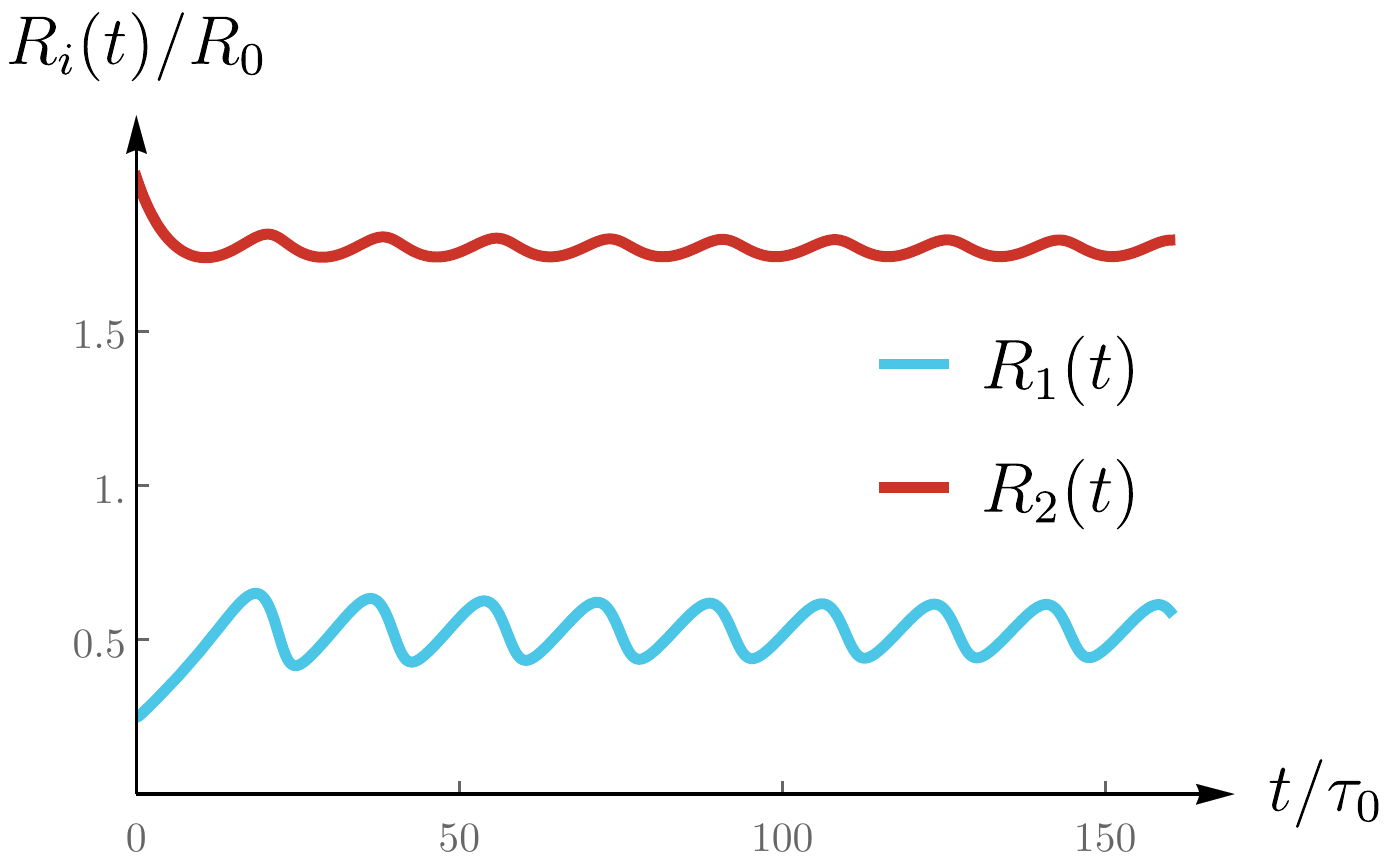}}
    \hfill
    \subfigure[\label{fig_osci2}]
    {\includegraphics[width=0.32\linewidth]{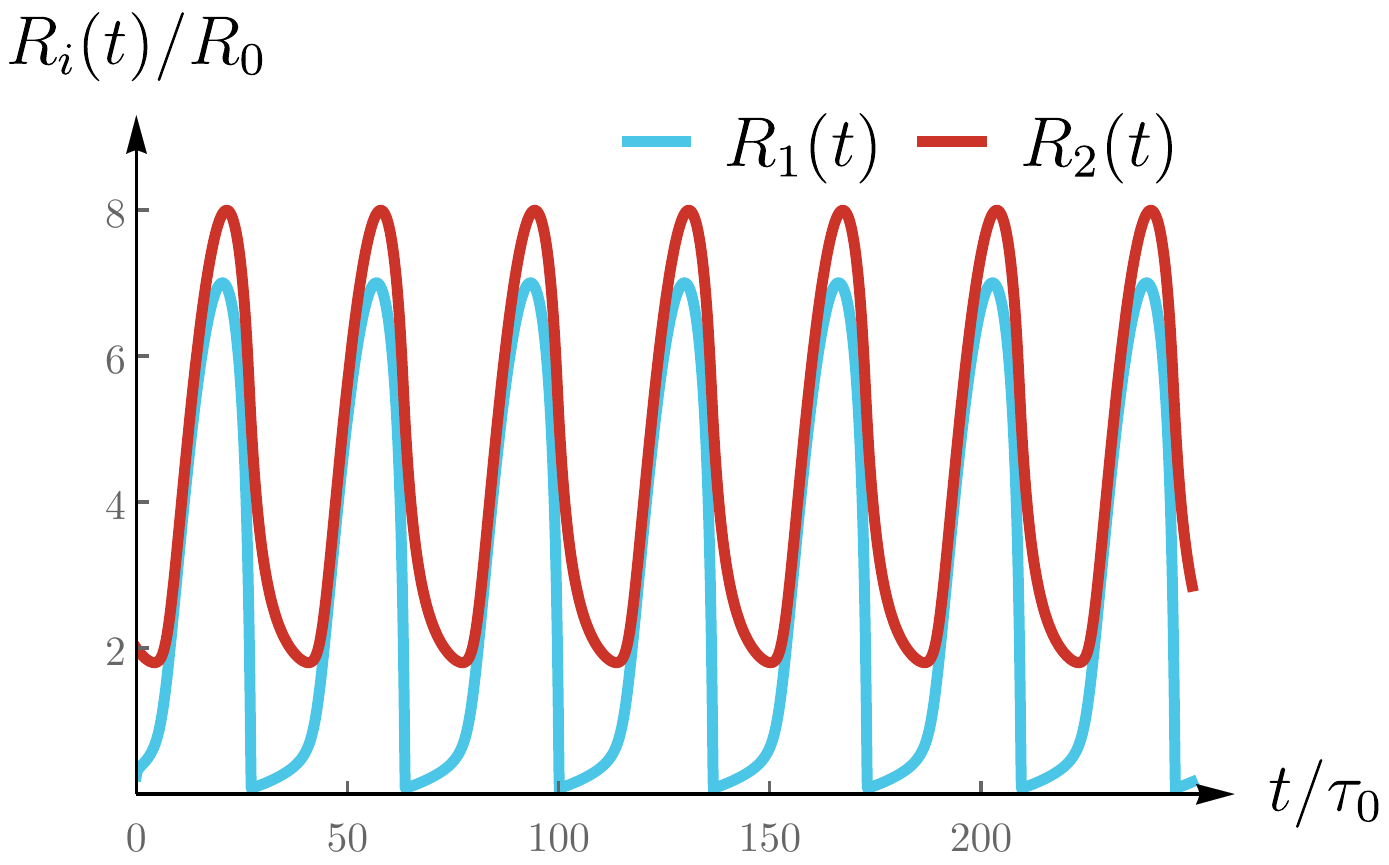}}
    \hfill
    \subfigure[\label{fig_osci3}]
    {\includegraphics[width=0.32\linewidth]{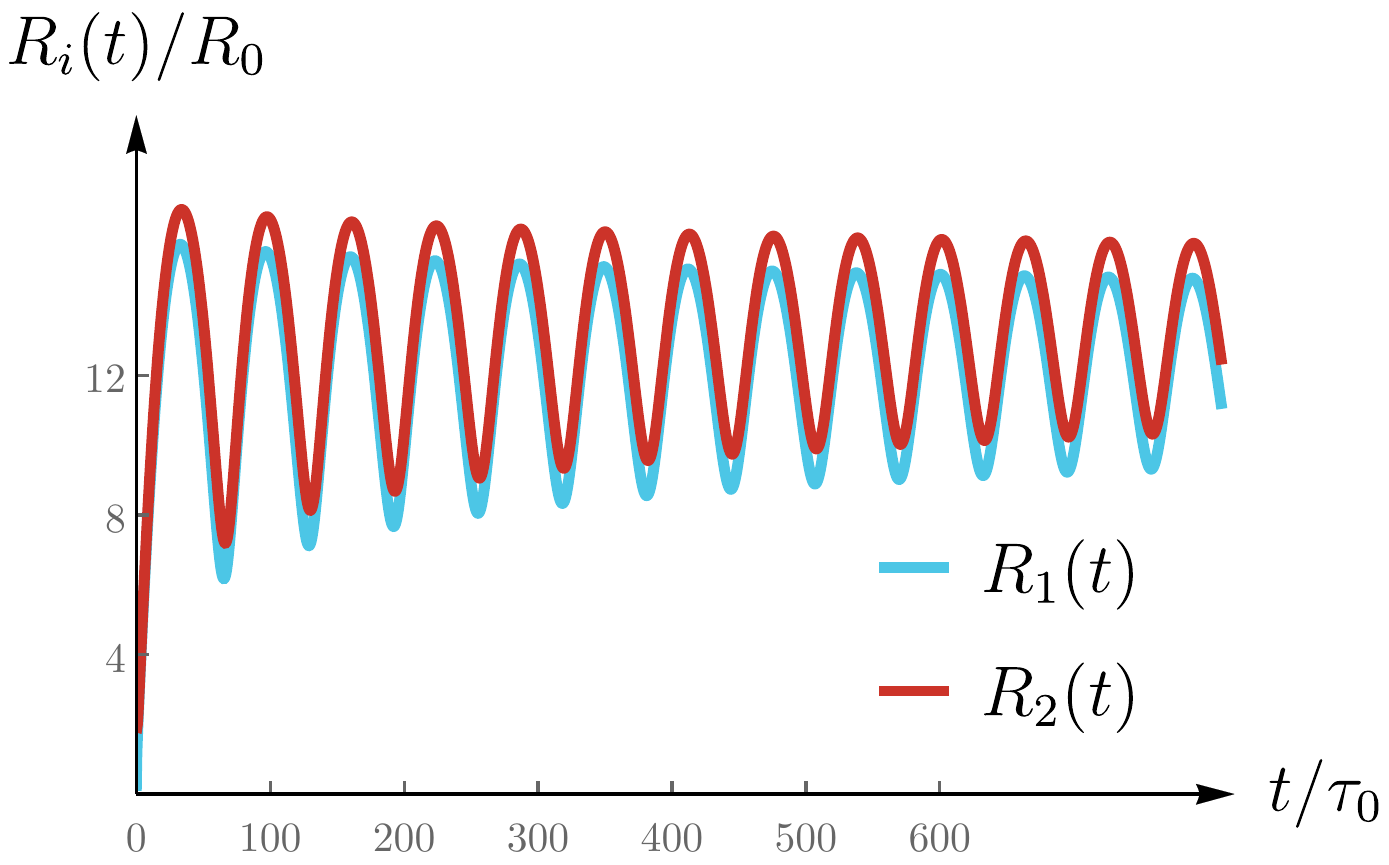}}
    \hfill \null
    \caption{(a) Eigenvalue $\mu$ (with ${\rm Im}(\mu)>0$) of the Jacobian of Eqs.~\eqref{eq_R1_adim} and \eqref{eq_R2_adim} evaluated at the fixed point, plotted in the complex plane as one varies $\hat \gamma_1$ at fixed value of $\hat\lambda\!=\!-2$ in Fig.~\ref{fig_stateDiagram}. As the control parameter $\hat \gamma_1$ is varied from $\hat \gamma_1\!=\!-0.2$ to $\hat \gamma_1\!=\!-2.5$, the eigenvalue follows the colored curve. Labelled dots along the curves indicate the value of the control parameter at these locations. The eigenvalue first crosses the imaginary axis at $\hat \gamma_1^-\simeq\!-0.22$, indicating a Hopf bifurcation and the entrance in the oscillating region. The eigenvalue crosses again the imaginary axis at $\hat \gamma_1^+\simeq\!-1.5$, which indicates the exit of the oscillating region. For a finite range of the control parameter inside the oscillating region, the eigenvalues are real (and the eigenvalue $\mu$ plotted here is the largest one in this case). For this range of parameters, the system however still reaches a stable limit cycle, indicating a strongly nonlinear behavior. The dynamics of the inner and outer radii ($R_1$ and $R_2$) at three different points in the oscillating regions (red dots) are displayed on the subfigures (b), (c) and (d). (b) Oscillations for $\hat\gamma_1\!=\!-0.218$. The amplitude of the oscillations is small and the spheroid thickness does not change significantly. (c) $\hat\gamma_1\!=\!-0.28$. Oscillations are strongly nonlinear and the spheroid thickness changes dramatically during one cycle. 
    (d) $\hat\gamma_1\!=\!-1.43$. The spheroid oscillations are almost sinusoidal. The proximity to the bifurcation point $\hat\gamma_1^+$ induces a  slow damping of the initial amplitude of the oscillations to the stationary amplitude.}
    \label{fig_oscillations_lambda_fixed}
\end{figure*}         
        
We discuss here the bifurcations between the different regions of the state diagram displayed in Fig.~\ref{fig_stateDiagram} and indicated by solid black lines (including the $x$ and $y$ axes). In the yellow region of the lower left quadrant the inner and outer radii reach a stable steady state $(R_1^*,R_2^*)$. As one goes from the lower left quadrant to the upper left quadrant, the fixed point existing in the lower quadrant moves progressively to larger values and is eventually sent to infinity as one crosses the $\hat\lambda=0$ axis. Going from the lower left quadrant to the lower right quadrant,  the value of the fixed point radius $R_1^*$ is sent progressively to~0 as $\hat\gamma_1$ increases. The fixed point radius eventually goes to negative $R_1^*$ values when the $\hat\gamma_1=0$ axis is crossed. Finally, the onset of spheroid oscillations (solid black curve between the yellow and red regions in the lower left quadrant) corresponds to the appearance of stable limit cycles through a Hopf bifurcation, as we discuss more in the following paragraphs.

We now focus our attention on the spheroid thickness oscillations (see Figs.~\ref{fig_oscillations} and \ref{fig_oscillations_lambda_fixed} for illustration) and on the characterization of the bifurcation between these oscillating states and the stable steady states. Spontaneous thickness oscillations are especially interesting in our model as they appear as a fine interplay between pumping and active electric effects (see state diagram, Fig.~\ref{fig_stateDiagram}). These electrohydraulic oscillations are indeed possible when flexoelectric effects spontaneously nucleate a lumen, while an outwards pumping of fluid acts for shrinking the cavity. If these two antagonistic processes are not balanced, which is the case when the system lies inside the red region in Fig.~\ref{fig_stateDiagram}, then the dynamics obeys a limit cycle around an unstable fixed point and thickness oscillations are predicted. Notice that the existence of such a limit cycle is not surprising, since, as discussed in the main text for a small lumen, the radii dynamical equations have a nonvanishing curl in phase space.

The bifurcation leading to limit cycles and oscillatory solutions in Fig.~\ref{fig_stateDiagram} is found to be a Hopf bifurcation. Indeed, let us define $\mu$ and $\mu'$ the two eigenvalues of the Jacobian associated to Eqs.~\eqref{eq_R1_adim} and \eqref{eq_R2_adim} evaluated at the fixed point of these equations. We also define the control parameter $\Gamma$ (for instance $\Gamma=\hat\gamma_1$ for a fixed value of $\hat\lambda <0$ in Fig.~\ref{fig_stateDiagram}), such that the system oscillates for $\Gamma_c^-\leq\Gamma\leq\Gamma_c^+$. For $\Gamma<\Gamma_c^-$ (or $\Gamma>\Gamma_c^+$), the eigenvalues of the system have a negative real part, such that the system reaches a stable steady-state at long time. Close to the boundary with the oscillating region, the eigenvalues are complex conjugated and the system spirals towards its fixed point. Precisely at the bifurcation ($\Gamma=\Gamma_c^\pm$), these eigenvalues cross the imaginary axis -- a signature of a Hopf bifurcation -- and the fixed point becomes unstable. The system is then driven to a limit cycle and displays periodic oscillations in time. We illustrate the crossing of the imaginary axis by the eigenvalues in Fig.~\ref{fig_eigenvalues}, for which we have chosen the control parameter to be $\Gamma=\hat\gamma_1$.

To illustrate the behavior of the system when varying $\hat\gamma_1$, we have plotted the dynamics of the spheroid and of the lumen radii as a function of time in Fig.~\ref{fig_oscillations_lambda_fixed}. Various kind of oscillations can be observed while $\hat\gamma_1$ is varied: small amplitude oscillations around the unstable fixed point as displayed in Fig.~\ref{fig_osci1}, large amplitude oscillations, strongly nonlinear and during which the lumen almost closes (see Fig.~\ref{fig_osci2}), or quasi-sinusoidal oscillations with a very thin thickness of the spheroid, as shown in Fig.~\ref{fig_osci3}. 

\section{Estimation of the parameter values}
\label{sec_estimations}

\begin{table}[t]
\setlength{\tabcolsep}{8pt}
\centering
    \begin{tabular}{ll||ll}
        \hline\hline
        \textbf{Parameters} & \textbf{Exp. values} & \textbf{Parameters} & \textbf{Estimations} \\
        \hline
        $\eta$ \cite{forgacs1998} & $10^{4}$~Pa$\cdot$s  & $\bar \kappa $ & $10^{3}$ A$\cdot$s$/$m$^{3}$  \\
        $\bar\eta$ \cite{montel2011} & $10^{9}$~Pa$\cdot$s & $\lambda_1$ & $10^8$ N$/$m$^{3}$  \\
        $\gamma_{1,2}$ \cite{forgacs1998} & $10^{-3}$~N/m & $\lambda_2, \, \lambda_3$ & $10^6$ N$/$m$^{2}/$V \\
        $\kappa^{-1}$~\cite{netti2000} & $10^{-13}$~m$^2$/Pa/s & $\lambda_4$ & $10^3$ N$/$m$^{2}$  \\
        $\Pi_{1,2}^{\rm ext}$  \cite{prather1968,*brace1977} & $10^3$ Pa & $\Lambda_1$ &  $10^0$ A$/$m$^{2}$  \\
        $v_{1,2}$ \cite{delarue2014} & $10^{-10}$~m/s & $\Lambda_2, \, \Lambda_3$ & $10^{-2}$ A$/$V/m \\
        $P^{\rm c}_{\rm h}$  \cite{montel2011} & $10^{3}$~Pa & $\Lambda_4$ & $10^{-5}$ A$/$m  \\
        $\zeta_0$ \cite{delarue2014} & $10^{3}$~Pa & $\nu_1$ & $10^1$ N/m/V  \\
        $\zeta_1$ \cite{delarue2014} & $-10^{-1}$ & $\nu_2$ & $10^8$ Pa$\cdot$s/m \\
        & & $\nu_3$ & $10^{0}$ N/m/V  \\
        & & $\nu_4$ & $10^8$ Pa$\cdot$s/m  \\
        & & $\Lambda_{1}^{\rm f}$, $\Lambda_{2}^{\rm f}$ & $10^{-11}$~m/Pa/s   \\
        & &$J_{{\rm p},1}$,  $J_{{\rm p},2}$ & $10^{-10}$~m/s \\
        & & $1-\phi$ & $10^{-2}$  \\ 
        & & $\nu_0$ & $1$  \\ 
        \hline\hline
    \end{tabular}
    \caption{Experimental values and references (left columns) and  estimated values (right columns) of the phenomenological parameters of the model appearing in the constitutive equations.}
    \label{table_estimationParam}
\end{table}

\begin{table}[t]
    \setlength{\tabcolsep}{8pt}
    \centering
    \begin{tabular}{llllllllll}
        \hline\hline
        \multirow{2}{*}{\textbf{Figure}} & \multicolumn{9}{c}{\textbf{Parameters values}}\\
        & $\delta_1$ & $\delta_2$ & $\chi$ & $\hat\gamma_0$  & $\hat\gamma_1$ & $\hat\gamma_2$ & $\hat \lambda$ & $\hat v_1$ & $\hat v_2$   \\
        \hline
        \ref{fig_LgrowthSgrowth} & 1 & -5 & 1 & 0  & -0.01 & 0.02 & 0 & 0.1 & 0.5   \\
        \ref{fig_LsteadySgrowth} & -1 & -0.83 & 1 & 0  & -$1.3\, 10^{-4}$ & $9.6\, 10^{-4}$ & 0 & $8.3\, 10^{-3}$ & $8.3\, 10^{-3}$   \\
        \ref{fig_ssWithLumen} & 1 & -0.1 & 1 & 0  & -0.05 & 0.15 & -5 & 0.1 & 0.1  \\
        \ref{fig_decay_to_ssMonolayer} & -1 & 33 & 1 & 0  & -0.03 & 0.037 & 0 & 0.33 & -0.13   \\
        \ref{fig_oscillations} & 1 & -1 & 1 & 0  & -0.3 & 0.4 & -4 & 0.1 & 0.1 \\
        \ref{fig_lumenClosure} & 1 & -0.9 & 10 & 0 & 0.4 & 0.4 & 0 & 0.01 & 0.3 \\
        \ref{fig_growth_to_monolayer} & 1 & 0.1 & 1 & 0  & -$4\, 10^{-3}$ & $6\, 10^{-3}$ & 0 & -0.05 & 0.05 \\
        \ref{fig_decay_to_monolayer_singular} & 1 & 1 & 1 & 0 & -0.4 & 0.6 & 0 & -2 & -1   \\
        \ref{fig_osci1}  & 1 & -1 & 1 & 0  & -0.218 & 0.318 & -2 & 0.1 & 0.1   \\
        \ref{fig_osci2}  & 1 & -1 & 1 & 0  & -0.28 & 0.38 & -2 & 0.1 & 0.1   \\
        \ref{fig_osci3}  & 1 & -1 & 1 & 0  & -1.43 & 1.53 & -2 & 0.1 & 0.1   \\
        \hline\hline
    \end{tabular}
    \caption{Dimensionless parameter values used for plotting the figures.}
    \label{table_parameters_plots}
\end{table}

Estimation of the different phenomenological parameters used in this coarse-grained spheroid model is essential in order to study the model in a biologically relevant regime and to simplify analytic computations. Some of the parameters (such as the cell shear and bulk viscosities, the surface tension etc.) have already been estimated in experiments. However, for most of the remaining phenomenological parameters such experimental values are not yet available, and we therefore used order-of-magnitude estimations to obtain them. Experimental and estimated  values are gathered in Table~\ref{table_estimationParam}. Most of the parameters of our model were already estimated in a previous work~\cite{sarkar2019}, and we detail here only the estimation of those that were not estimated, namely the coupling of the isotropic and anisotropic stresses to the velocity difference $\nu_2$ and $\nu_4$, the coupling of curvature and pumping parameter $\lambda_4$, and the active flexoelectricity parameter $\Lambda_4$.

The coefficient $\nu_2$ is estimated by assuming that $\nu_2 p_\alpha (v^{\rm c}_\alpha-v^{\rm f}_\alpha)$ is the stress due to the hydraulic friction force density $\kappa(v^{\rm c}_\alpha-v^{\rm f}_\alpha)$, such that we obtain $\nu_2 \sim \ell \kappa \simeq 10^{8}~\text{Pa$\cdot$s/m}$ where $\ell\sim 10~\mu$m is the typical size of a cell. The coefficient $\nu_4$ is estimated by assuming that its contribution to the anisotropic cell stress is of the same order of magnitude as the viscous stress in the cleft between cells, that is $\tilde \sigma_{xz}\sim \eta^{\rm f} \delta v_x/w$ where the interstitial fluid channels width is $w\sim 50$~nm and we have introduced the viscosity of the interstitial fluid $\eta^{\rm f}\sim100$~mPa$\cdot$s. We then obtain the estimate $\nu_4\sim \eta^{\rm f}\ell/w^2 \simeq 10^8~\text{Pa$\cdot$s/m}$.

The coupling of curvature and pumping parameter $\lambda_4$ is estimated by noticing that for in a flat geometry, a  cell produces a flow $v\sim\kappa^{-1}\lambda_1$ due to pumping. Bending this cell implies an extra flow $v'\sim v \delta A/A$ where $A$ is the cell area and $\delta A/A \sim \ell C$ (with $C$ the curvature of the bent cell) is the extra area due to bending and which contributes to the extra flow. This extra pumping due to curvature produces a flow which is by definition of the order $v'\sim \kappa^{-1} C \lambda_4$. We therefore obtain $\lambda_4 \sim \lambda_1 \ell \sim 10^3~\text{N$\cdot$m}^{-2}$. The flexoelectricity parameter $\Lambda_4$ is obtained using the a similar argument and we obtain $\Lambda_4 \sim \Lambda_1 \ell \sim 10^{-5}~\text{A$\cdot$m}^{-1}$.